\documentclass[lettersize,journal]{IEEEtran}
\usepackage{amsmath,amsfonts}
\usepackage{algorithmic}
\usepackage{algorithm, color}
\usepackage{array}
\usepackage[font=small]{caption}
\usepackage{textcomp}
\usepackage{stfloats}
\usepackage{url}
\usepackage{verbatim}
\usepackage{graphicx}
\usepackage{cite}
\usepackage[T1]{fontenc}
\usepackage{theorem}
\usepackage{amssymb}
\usepackage{balance}
\usepackage{bm}
\usepackage{subfigure}
\allowdisplaybreaks[4]

\newtheorem{lemma}{Lemma}\theoremheaderfont{\normalfont\bfseries}\newtheorem{Remark}{Remark}

\begin{document}

\title{Joint Active and Passive Beamforming Design for IRS-aided MIMO ISAC Based on Sensing Mutual Information}

\author{
Jin Li,~Gui Zhou,~\IEEEmembership{Member,~IEEE},~Tantao Gong,~Nan Liu, and Rui Zhang,~\IEEEmembership{Fellow,~IEEE}
\thanks{
	This article was presented in part at the IEEE PIMRC Workshops 2024.
	
	J. Li, T. Gong and N. Liu are with the National Mobile Communications Research Laboratory, Southeast University, Nanjing 211189, China. (e-mail: lijin, gongtantao, nanliu@seu.edu.cn). G. Zhou is with the Institute for Digital Communications, Friedrich-Alexander-University Erlangen-N\"{u}rnberg (FAU), 91054 Erlangen, Germany (email: gui.zhou@fau.de). R. Zhang is with School of Science and Engineering, Shenzhen Research Institute of Big Data, The Chinese University of Hong Kong, Shenzhen, Guangdong 518172, China (e-mail: rzhang@cuhk.edu.cn). He is also with the Department of Electrical and Computer Engineering, National University of Singapore, Singapore 117583 (e-mail: elezhang@nus.edu.sg).}

}

%
%

\maketitle

\begin{abstract}
In this paper, we investigate the intelligent reflecting surface (IRS)/reconfigurable intelligent  surface (RIS)-aided integrated sensing and communication (ISAC) system based on sensing mutual information (MI).
Specifically, the base station (BS) perceives the sensing target via the reflected sensing signal by the IRS,  while communicating  with the users simultaneously.  
Our aim is to maximize the sensing MI, subject to the quality of service (QoS) constraints for all communication users, the transmit  power constraint at the BS, and the unit-modulus constraint on the IRS's passive reflection.
We solve  this problem under two cases: one simplified case assuming  a line-of-sight (LoS) channel between the BS and IRS and no clutter interference to sensing, and the other  generalized case considering the   Rician fading channel  of the BS-IRS link and the presence of clutter interference to sensing.
For the first case, we prove that the dedicated sensing beamformer is unnecessary for improving sensing MI and develop a low-complexity iterative algorithm to jointly optimize the BS and IRS active/passive beamformers.
Then, for the second case, 
we propose an alternative  iterative algorithm, which can also be applied to the first  case, 
to solve the beamforming design problem under the general setup. 
Numerical results are provided to validate the performance of the proposed algorithms, as compared to various  benchmark schemes.

\end{abstract}


\vspace{-5mm}
\section{Introduction}
The next generation (6G)  wireless networks with  both  high-accuracy sensing and high-quality communication are expected to support various  applications more capably, such as environmental monitoring, smart home, vehicle to everything (V2X) communication  and industrial internet of things (IoT) \cite{isacoverview,fantcom,cui,perceptive}.      
In the existing literature, there are two general approaches  to co-design sensing and communication systems: radar communication co-existence (RCC)  and integrated sensing and communication (ISAC)  \cite{fantcom}. 
RCC systems allocate the spectrum resource for sensing and communication appropriately to achieve their  co-existence and reduce mutual interference \cite{rccoverview}. By comparison, ISAC systems, also called dual-functional radar and communication (DFRC) or joint radar and communication (JRC) systems, can obtain integration and coordination gains  by forming  a unified system with shared spectrum, energy   and hardware \cite{fantcom}, \cite{cui}. 
 On the other hand, multiple-input multiple-output (MIMO), which can greatly improve the sensing performance compared with traditional phase-array radar and effectively compensate  the high path-loss and attenuation of millimeter wave (mm-Wave) communications \cite{lijian}, \cite{mmwave}, has been  widely applied in ISAC systems \cite{CRB}, \cite{fantwc}. 
 However, the energy consumption and hardware complexity of MIMO systems  can go up drastically  with the increasing  number of antennas equipped due to their associated radio frequency (RF) chains.     
 Thus, intelligent reflecting surface (IRS)/reconfigurable intelligent  surface (RIS) has been recognized as a promising technology for MIMO in  6G, which can offer  the same performance  advantages of conventional MIMO but is more cost-effective due to the use of passive reflection elements only \cite{irscom}. 
In a non-favorable environment with dense buildings, although the direct sensing/communication link between the base station (BS)  and the sensing target/users is blocked by obstacles,  IRS can still assist in the sensing or communication task via the reflected link \cite{irscom,multicast,songCRB}. 
Therefore, IRS can be used to improve the performance of MIMO ISAC  systems. 
 

 Based on this background,
 extensive works studied IRS-aided MIMO sensing \cite{shaoxiaodan}, \cite{wangpeilan} and MIMO ISAC systems,
 where the sensing performance is measured by different metrics, such as  signal-to-interference-plus-noise-ratio (SINR), detection probability, beampattern, or Cramér-Rao bound (CRB) \cite{lijian}.
 In \cite{huanglei,liurang,sensingsinr-3,huamengsinr,lr-sinr}, the sensing SINR  was adopted to design the ISAC beamforming of the BS and the passive beamforming of the IRS while the communication performance is satisfied. 
 The authors in \cite{dtprobability} and \cite{dtprobability-2} studied  IRS-aided ISAC systems based on the target detection probability  and  the weighted communication mean square error (MSE).
 Song  \emph{et al.} considered the CRB of non-line-of-sight (NLoS) sensing link  in the IRS-aided ISAC system and derived the  estimation CRB \cite{songisacCRB}. 
 Moreover,
 beamforming design for IRS-aided sensing  based on beampattern metric was also studied  in \cite{songwcnc,bp-1,benchmark-bp,hua-bp}.
However, the sensing metrics above may have some limitations. For instance, CRB is not tight in low SNR, detection probability has very complicated mathematical expression and relies on specific detector such as Neyman-Pearson detector, beampattern is a transmit metric without considering echo signal, and radar SNR maximization  needs to design extra received filter.

In light of these limitations, mutual information (MI) between the echo signal and the target response matrix has been proposed as another important metric for evaluating the overall sensing performance.
MI can be also interpreted as the reduction of uncertainty about the target response matrix given the echo signal and transmit signal.
The authors in \cite{xielei} have shown that the MI is the maximum information 
about the target response matrix extracted from the  echo signal by Bayesian estimator. 
In addition,
other  works  have exhibited
 the benefits of waveform design based on MI. 
 In radar systems, when the target response matrix is drawn according to Gaussian distribution, waveform design with higher sensing MI can achieve better radar classification and estimation performance \cite{mi}. 
 Furthermore, compared to phase array radar, MIMO radar can co-exist with communication systems more efficiently by using waveform design based on sensing MI \cite{tangboradar}. 
 Additionally,  sensing MI can provide a universal lower bound  
 for some sensing distortion metrics, such as detection probability and MSE \cite{fanisit}. 
 In \cite{jin}, it was shown that MIMO ISAC beamforming scheme based on sensing MI metric can suppress the clutter interference effectively and
 attain  better transmit beampattern and root mean square error (RMSE) of angle estimation compared with MIMO ISAC beamforming schemes based on CRB and beampattern metrics. 
 Several recent works also adopted sensing MI metric to measure the sensing performance in IRS-aided MIMO ISAC systems. 
 In \cite{zhl,irs-mi}, an iterative algorithm based on manifold optimization have been proposed to maximize the weighted sum of communication rate and  sensing MI, where the overall communication and sensing performance can be ensured.
 Moreover,  due to the quartic function of IRS reflection coefficients, only heuristic algorithms or gradient descent algorithms based on manifold optimization are applicable \cite{irs-mi}.


Different from the above works, our paper investigates an IRS-aided MIMO ISAC system based on MI, where the direct sensing link from the BS to the sensing target is blocked by obstacles. To overcome this, an IRS is deployed to create a reflected sensing link to bypass the obstacles \cite{active-jin}. 
We consider signal-dependent clutter interference and a dedicated sensing beamformer, and aim to jointly design the transmit beamformer at the BS and the reflection coefficients at the IRS to maximize the sensing MI, while ensuring communication fairness.
Unfortunately,
existing heuristic or gradient descent methods in \cite{zhl,irs-mi} may result in  poor performance on sensing MI  and are applicable only to  the weighted sum of communication rate and sensing MI, but not to our more general  scenario considering clutter interference and communication fairness.
In addition, the impact of the dedicated sensing beamformer on sensing MI is also not studied yet.
Hence, in this paper, we study an MI-based IRS-aided MIMO ISAC system with clutter interference, communication fairness, and a dedicated sensing beamformer.  
As an extension  of our previous conference work, we consider a more general case with a Rician BS-IRS channel and clutter interference to sensing. We derive a more general conclusion regarding the optimality of the dedicated sensing beamformer and the communication beamformer under certain conditions, establish an upper bound for sensing MI to assess the algorithm's performance  with a point sensing target, and present additional numerical results to support our conclusions.
The main contributions of this  paper are summarized as follows:


$\bullet$    
In this paper, we  aim to maximize the sensing MI, subject to the QoS constraints for all communication users, the transmit power constraint at the BS and the unit-modulus constraint on the IRS's passive reflection coefficients.
The BS transmit beamformer and IRS reflection coefficients are jointly optimized.
More specifically,  in order to improve algorithm performance and study the impact of dedicated beamformer,
we consider two cases: one simplified case considering a line-of-sight (LoS)  BS-IRS channel and no clutter interference to sensing, and the other generalized case with a Rician fading NLoS   BS-IRS channel and the presence of clutter interference to sensing.

$\bullet$ 
In the simplified case,
the nonconvex quartic sensing MI is firstly simplified by exploring the rank-one property of the LoS BS-IRS channel.
Then, by using alternating optimization (AO), 
we prove that for the BS beamformer subproblem, the dedicated sensing beamformer is unnecessary for improving sensing MI. 
Besides, we propose a low-complexity algorithm to solve the formulated  problem, where semidefinite relaxation (SDR),  eigenvalue decomposition (EVD), and convex-concave procedure (CCP) methods are adopted to solve the problem efficiently, where close  MI performance can be achieved compared to the upper bound of MI.
However, this low-complexity algorithm can only be applied to the simplified case.

$\bullet$
In the generalized case, we propose a more general algorithm, which can also be applied to the simplified case. 
To cope with the intractable nonconvex quartic sensing MI,
the majorization-minimization (MM) method is adopted to approximate the quartic objective function as a biconcave minorizer. Then, an alternative  iterative algorithm is proposed for this general setup based on  various methods, such as  matrix transformations,  SDR and  Gaussian randomization.

$\bullet$
Numerical results demonstrate that, in the simplified case, the proposed low-complexity iterative algorithm offers both lower computational complexity  and better performance compared to the alternative iterative algorithm designed for the generalized case. Additionally, in the simplified case, 
the dedicated sensing beamformer is unnecessary for improving sensing MI.
Moreover, in the scenario with an LoS BS-IRS channel, the beampattern performance may be worse compared to the scenario with a Rician fading channel for the BS-IRS link, as the BS is unable to sense all directions of the extended sensing target in the former case. Finally, compared to other benchmarks, our proposed algorithms achieve superior performance in terms of  both sensing MI and beampattern.

The rest of this paper is organized as follows.
Section \uppercase\expandafter{\romannumeral2} introduces the system model and derives the closed-form  expression of sensing MI.
The optimization problems  for the simplified case  
and the generalized case  
are respectively investigated in Sections \uppercase\expandafter{\romannumeral3} and \uppercase\expandafter{\romannumeral4}.
Numerical results and conclusions are provided in Sections \uppercase\expandafter{\romannumeral5} and  \uppercase\expandafter{\romannumeral6}, respectively.

\textbf{Notations:}
The following mathematical notations are adopted throughout this paper.
Scalars, vectors, and matrices are denoted by normal font, boldface lowercase letters,
and boldface uppercase letters, respectively.
The symbols $\mathbf{A}^T$, $\mathbf{A}^*$, $\mathbf{A}^H$, and $\mathbf{A}^{-1}$ represent the transpose, conjugate, hermitian, and inverse operations of matrix $\mathbf{A}$, respectively.
The $l_2$ norm of vector $\mathbf{a}$ is denoted by $\Vert \mathbf{a} \Vert$, and the 
Frobenius norm of matrix $\mathbf{A}$ is denoted by ${\Vert \mathbf{A} \Vert}_F$.
The symbols $\det (\mathbf{A}), \text{Tr} (\mathbf{A})$, and $\text{Re} (\mathbf{A})$ represents the determinant, trace, and real part of a matrix, respectively.
Complex field,  real field, and the imaginary unit  are denoted as $\mathbb{C}$, $\mathbb{R}$, and $j = \sqrt{-1}$, respectively.
$\vert a \vert$ and  $\angle a$ denote the modulus and angle of complex number $a$, respectively.
$\lambda_{min}(\mathbf{A})$ and $\lambda_{max}(\mathbf{A})$ represent the minimum and maximum eigenvalues of matrix $\mathbf{A}$. 
$[\mathbf{A}]_{p:q,m:n}$ denotes a matrix with elements being the $p$-th to  the $q$-th rows and  the $m$-th to  the $n$-th columns of matrix $\mathbf{A}$, and  $[\mathbf{a}]_{m}$ denotes the $m$-th element of vector $\mathbf{a}$.
The Kronecker product and Hadamard product between matrices $\mathbf{X}$ and $\mathbf{Y}$ are respectively denoted by 
$\mathbf{X} \otimes \mathbf{Y} $ and $\mathbf{X} \circ \mathbf{Y} $.
$\mathrm{diag}(\mathbf{a})$ represents the diagonal matrix with the main diagonal elements being the entries of vector $\mathbf{a}$, and   $\mathrm{Diag}(\mathbf{A})$ denotes the vector with entries from the main diagonal elements of matrix $\mathbf{A}$.
The vectorization of matrix $\mathbf{A}$ is represented by $\text{vec} (\mathbf{A})$, and 
$\text{unvec}_{N,N}(\mathbf{a})$ denotes undo the vectorization, i.e., 
matrixizing a vector $\mathbf{a}$ into an $N \times N$ matrix.
Complex Gaussian distribution is denoted by $\mathcal{CN}$,
$\mathbf{I}_a$ denotes the identity matrix of size $a$, $\mathbf{0}_{a \times b}$ denotes the  matrix of size $a \times b$ with every element being 0, 
and $\mathbf{1}_{a}$ represents the column vector of size $a$ with every element being 1.

\vspace{-3mm}
\section{System Model}
\begin{figure} 
	\centering 
	\includegraphics[width=2.8in,height=1.8in]{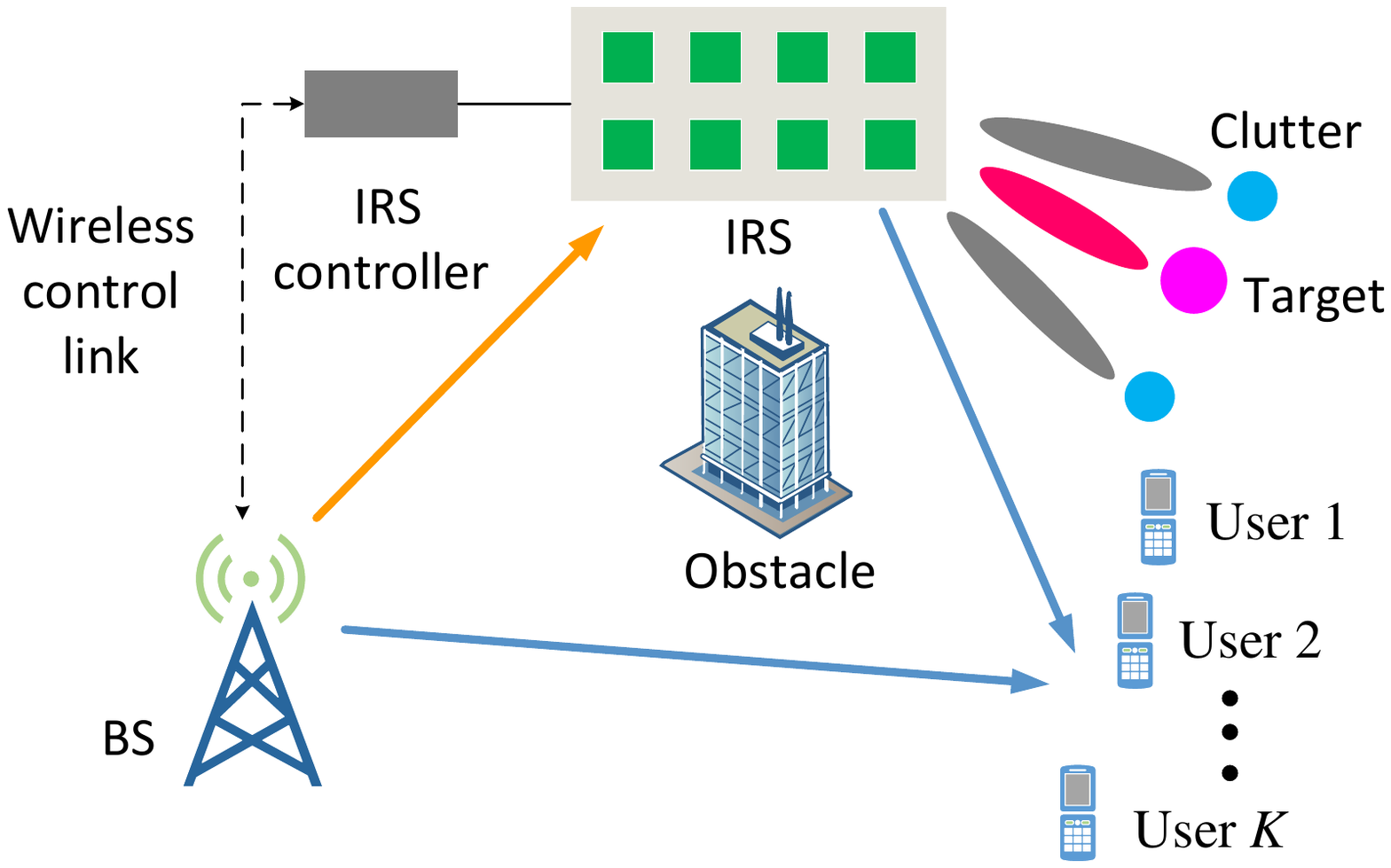}
	\caption{An IRS-aided ISAC system.}
	\label{system} 
\end{figure}

As shown in Fig. \ref{system}, we consider an IRS-aided ISAC system, where a MIMO DFRC BS equipped with an $N$-antenna uniform linear array (ULA) senses a radar target and transmits information to $K$ single-antenna communication users simultaneously, and an IRS equipped with a $M$-reflecting-element ULA is deployed to assist in both sensing and communication. 
We assume that the direct sensing link between the BS and the radar target is severely  blocked by  obstacles\footnote{In environments with dense high-rise buildings or many indoor walls, direct sensing links are often blocked, where only the reflected link created by the IRS is useful for wireless sensing and human-activity recognition \cite{xiaodanwc}.  }, and as a result, the BS working as a monostatic radar  can only sense the radar target via the reflected sensing link created by the IRS.

To utilize full degrees-of-freedoms (DoFs) of the transmit signal to improve radar beampattern, individual communication and sensing waveforms are applied, where 
transmit waveform can achieve the maximum DoFs, i.e., the number of antennas \cite{CRB}, \cite{liuxiang}.
Thus, the transmit DFRC signal at the BS is given by 
\begin{align}
	\mathbf{X} = \mathbf{W} \mathbf{S}  
= \begin{bmatrix} \mathbf{W}_{\mathrm{C}}  &  \mathbf{W}_{\mathrm{R}}  \end{bmatrix}
\begin{bmatrix}
	\mathbf{S}_{\mathrm{C}}  \\
	\mathbf{S}_{\mathrm{R}}
\end{bmatrix} \in \mathbb{C}^{N \times L}, \nonumber
\end{align}
where 
 $\mathbf{W} \in \mathbb{C}^{N \times (K+N)} $ is the dual-functional beamforming matrix at the BS with $\mathbf{W}_{\mathrm{C}} = [\mathbf{w}_1, \mathbf{w}_2, \cdots, \mathbf{w}_K ] \in \mathbb{C}^{N \times K}$ and $\mathbf{W}_{\mathrm{R}} \in \mathbb{C}^{N \times N}$ being the beamforming matrices for communication and sensing, respectively,
 and $ \mathbf{S} 
\in \mathbb{C}^{ (K+N) \times L}$ is the signal  matrix with 
$\mathbf{S}_{\mathrm{C}}  \in \mathbb{C}^{K \times L}$ and $\mathbf{S}_{\mathrm{R}} \in \mathbb{C}^{N \times L}$ being $K$ communication signals  intended for $K$ users  and dedicated probing signals for target sensing, respectively.
Here, $L$ is the length of communication time slots as well as  the number of radar fast-time snapshots. Throughout this paper, we assume that $K \leq N \leq L$ for the purpose of exposition  \cite{CRB}. Additionally, we assume that each entry in $\mathbf{S}_{\mathrm{C}}$ is i.i.d,  and follows the complex Gaussian distribution with zero mean and unit variance, and $\mathbf{S}_{\mathrm{R}}$ is generated by pseudo random coding and orthogonal to  $\mathbf{S}_{\mathrm{C}}$ \cite{CRB}, \cite{liuxiang}.
Therefore, according to the law of large numbers, if $L$ is asymptotically large,
 we have $\frac{1}{L} \mathbf{S}_{\mathrm{R}} \mathbf{S}_{\mathrm{R}}^H  \approx \mathbf{I}_{N}$, $\frac{1}{L} \mathbf{S}_{\mathrm{C}} \mathbf{S}_{\mathrm{C}}^H  \approx \mathbf{I}_{K}$, and $\frac{1}{L} \mathbf{S}_{\mathrm{C}} \mathbf{S}_{\mathrm{R}}^H  = \mathbf{0}_{K \times N}$.
Based on the assumptions above, the following condition holds:
\begin{align}
	\frac{1}{L} \mathbf{S} \mathbf{S}^H  \approx \mathbf{I}_{K+N}. \label{approx}
\end{align}	

The reflection coefficients matrix of the IRS is represented by $\mathbf{E} = \mathrm{diag}( e_1, e_2, \cdots, e_M) \in \mathbb{C}^{M \times M}$, where $|e_m|^2 = 1, m = 1, \cdots, M$. In addition, the channels between the BS and the IRS, the IRS and the $k$-th user, and the BS and the $k$-th user are denoted by $\mathbf{H}_{\mathrm{BI}} \in \mathbb{C}^{M \times N}$, $\mathbf{h}_{\mathrm{IU},k} \in \mathbb{C}^{M \times 1}$, and 
$\mathbf{h}_{\mathrm{BU},k} \in \mathbb{C}^{N \times 1}$, respectively. 
It is assumed that perfect channel state information (CSI)\footnote{The BS-IRS CSI can be obtained by a full-duplex BS using Khatri-Rao factorization and ambiguity identification \cite{zgchannel}, or an anchor-aided channel estimation technique \cite{anchor}. Then, the BS-users CSI can be acquired by least square and MMSE estimations with known pilot. } for the above communication links  is known at the BS, such that  the BS designs the beamforming matrix and the reflection coefficients matrix of the IRS jointly, while the latter is sent  to the IRS controller via a wireless control link as shown in Fig. \ref{system} for implementation  \cite{irscom}.

\vspace{-4mm}
\subsection{Communication Model}
The received signal at the $k$-th user in the $l$-th time slot can be written as 
\begin{align}
   y_{k,l} = ( \mathbf{h}_{\mathrm{BU},k}^H + \mathbf{h}_{\mathrm{IU},k}^H \mathbf{E} \mathbf{H}_{\mathrm{BI}} ) \mathbf{x}_l + n_{k,l}, 
\end{align}
where $n_{k,l}$ is the additive white Gaussian noise (AWGN) following distribution $n_{k,l} \sim \mathcal{CN} (0,\sigma_{k}^2 )$.

Then, the achievable rate of the $k$-th user is given by 
\begin{align}
	R_k 
	= \log_2 \Big(1+ \frac{{\lvert  \mathbf{h}_k^H \mathbf{w}_k \rvert}^2}{ \sum_{i=1, i\neq k }^{K}  {\lvert \mathbf{h}_k^H \mathbf{w}_i \rvert}^2   + {\Vert \mathbf{h}_k^H \mathbf{W}_{\mathrm{R}} \Vert}_2^2 + \sigma_{k}^2} \Big),  \label{com-sinr}
\end{align} 
where $\mathbf{h}_k^H = \mathbf{h}_{\mathrm{BU},k}^H + \mathbf{h}_{\mathrm{IU},k}^H  \mathbf{E} \mathbf{H}_{\mathrm{BI}}$.
Note that while the dedicated sensing beamformer improves sensing performance, it also introduces communication interference. Therefore, optimizing the dedicated sensing beamformer is essential to maximize overall system performance.


\vspace{-4mm} 
\subsection{Radar/Sensing  Model}
In the following, as shown in Fig. \ref{system}, the clutter interference caused by scatterers around the radar target is considered \cite{tangrelative}.
Thus, the echo signal received at the BS can be written as 
\begin{align}
	\mathbf{Y}_R = \mathbf{H}_{\mathrm{BI}}^H \mathbf{E}^H (\mathbf{G}_R + \mathbf{G}_C) \mathbf{E} \mathbf{H}_{\mathrm{BI}} \mathbf{X} + \mathbf{Z}_R,
\end{align}
where $\mathbf{G}_R \in \mathbb{C}^{M \times M}$ is the target response matrix,  $\mathbf{G}_C \in \mathbb{C}^{M \times M}$ is the interference response matrix, and $\mathbf{Z}_R$ is the AWGN matrix with each entry of $\mathbf{Z}_R$ being i.i.d, and following the complex Gaussian distribution with zero mean and variance $\sigma_{R}^2$, i.e., $\text{vec} (\mathbf{Z}_R^H) \sim \mathcal{CN} (\mathbf{0},\sigma_{R}^2 \mathbf{I}_{NL} ) $.

By vectorizing $\mathbf{Y}_R^H$, we have 
\begin{align}
	\mathbf{y}_R &= \text{vec}  (\mathbf{Y}_R^H) \nonumber \\
	             &= \big[   (\mathbf{H}_{\mathrm{BI}}^T  \mathbf{E}^T) \otimes (\mathbf{X}^H  \mathbf{H}_{\mathrm{BI}}^H  \mathbf{E}^H) \big] ( \mathbf{g}_R   + \mathbf{g}_C) + \mathbf{z}_R,
\end{align} 
where $ \mathbf{g}_R = \text{vec} (\mathbf{G}_R^H),  \mathbf{g}_C = \text{vec} (\mathbf{G}_C^H)$ and $\mathbf{z}_R = \text{vec} (\mathbf{Z}_R^H)$. We assume that $\mathbf{g}_R$ and $\mathbf{g}_C$ follow the complex Gaussian distribution, i.e., $ \mathbf{g}_R \sim \mathcal{CN} (\mathbf{0},\mathbf{R}_R )$ and $ \mathbf{g}_C \sim \mathcal{CN} (\mathbf{0},\mathbf{R}_C )$.  
We also assume that $\mathbf{g}_R$, $\mathbf{g}_C$, and   $\mathbf{z}_R$ are independent to each other.

In order to calculate the covariance matrix  $\mathbf{R}_R$, we first introduce the model of target response matrix $\mathbf{G}_R$. We assume that IRS works as a monostatic colocated MIMO radar \cite{lijian}, \cite{songCRB}, i.e., the angle of arrival (AOA) is equal to the angle of departure (AOD). 
Hence, according to the Swerling \uppercase\expandafter{\romannumeral1} target models, the target response matrix $\mathbf{G}_R$ can be expressed as   \cite{tangboradar}, \cite{swerling}
\begin{align}
	\mathbf{G}_R = \sum_{p=1}^{N_r} \alpha_p^r \mathbf{b} (\theta_p^r)  \mathbf{b}^H (\theta_p^r), \label{G_R}
\end{align}
where $N_r$ is the number of scatterers, $\theta_p^r$ is the AOA/AOD of the $p$-th scatterer relative to the IRS, $\alpha_p^r$ is the complex amplitude of the $p$-th scatterer containing the round-trip pathloss and the radar cross section of the scatterer with $\alpha_p^r \sim \mathcal{CN} (0, \beta_p^2), \beta_p^2 = \mathbb{E} (\alpha_p^r \alpha_p^{r,*}  ) $, and $\mathbf{b} (\theta) \in \mathbb{C}^{M \times 1}$ is the steering vector of the IRS. 
Here, the radar target is considered as an extended target,  where multiple point scatterers induce  different propagation paths. The BS has no prior knowledge of the number of propagation paths or the angles associated with the point scatterers.
Hence, we aim to estimate the target response matrix and 
 extract the desired parameters from $\mathbf{G}_R$.

With the target response matrix \eqref{G_R}, the covariance matrix $\mathbf{R}_R$ is given by
\begin{align}
	\mathbf{R}_R &= \mathbb{E} ( \mathbf{g}_R \mathbf{g}_R^H ) \nonumber\\
	&=\sum_{p=1}^{N_r} \beta_p^2 (  \mathbf{b}^* (\theta_p^r)    \otimes \mathbf{b} (\theta_p^r) ) (  \mathbf{b}^* (\theta_p^r)    \otimes \mathbf{b} (\theta_p^r) )^H.  \label{R_R}
\end{align}

We consider that the clutter interference 
 is resulted from many scatterers around the radar target. Similar to \eqref{G_R}, \eqref{R_R}, the interference response matrix $\mathbf{G}_C$ and its covariance matrix $\mathbf{R}_C$ can be respectively written as
\begin{align}
	\mathbf{G}_C &= \sum_{q=1}^{N_c} \alpha_q^c \mathbf{b} (\theta_q^c)  \mathbf{b}^H (\theta_q^c), \label{G_C} \\
	\mathbf{R}_C &= \mathbb{E} ( \mathbf{g}_C \mathbf{g}_C^H ) \nonumber \\
	&=\sum_{q=1}^{N_c} \gamma_q^2 (  \mathbf{b}^* (\theta_q^c)    \otimes \mathbf{b} (\theta_q^c) ) (  \mathbf{b}^* (\theta_q^c)    \otimes \mathbf{b} (\theta_q^c) )^H,  \label{R_C}
\end{align}
where   $\gamma_q^2 = \mathbb{E} (\alpha_q^c \alpha_q^{c,*}  ) $, $N_c$ is the number of interfering scatterers, $\alpha_q^c$ is the complex amplitude of the $q$-th interfering scatterer, and $\theta_q^c$ is the AOA/AOD of the $q$-th interfering scatterer relative to the IRS. 
 We assume a prior knowledge of the second-order statistics of both the sensing target and the clutter, which can be obtained by previous scans or a dynamic environmental database, such as a geographical information system (GIS), the National Land Cover Database (NLCD) and meteorological data sources \cite{cognitive,knowledge,tangrelative}. 

To evaluate the sensing performance, the MI between the object parameter vector $\mathbf{g}_R$ containing the desired parameters and  the measurement vector $\mathbf{y}_R$ is adopted as the sensing performance metric \cite{mi}. 
Then, according to the aforementioned assumptions, the channel between the BS and the IRS, $\mathbf{H}_{\mathrm{BI}}$, the reflection coefficients matrix, $\mathbf{E}$,  and the transmit signal at the BS, $\mathbf{X}$, are known at the DFRC BS.
Hence, given $\mathbf{H}_{\mathrm{BI}}$, $\mathbf{E}$, and $\mathbf{X}$, the MI between $\mathbf{y}_R$ and $\mathbf{g}_R$ can be written as \cite{tangboradar} 
\begin{align}
&\quad \mathcal{I} (\mathbf{y}_R ; \mathbf{g}_R \vert \mathbf{X}, \mathbf{H}_{\mathrm{BI}}, \mathbf{E}  ) \nonumber \\
&= \mathcal{H} (\mathbf{y}_R  \vert \mathbf{X}, \mathbf{H}_{\mathrm{BI}}, \mathbf{E}  ) - \mathcal{H} (\mathbf{y}_R  \vert \mathbf{g}_R, \mathbf{X}, \mathbf{H}_{\mathrm{BI}}, \mathbf{E}  ) \nonumber \\
&= - \int p(\mathbf{y}_R \vert  \mathbf{X}, \mathbf{H}_{\mathrm{BI}}, \mathbf{E})	\log \; p(\mathbf{y}_R \vert  \mathbf{X}, \mathbf{H}_{\mathrm{BI}}, \mathbf{E})   
\nonumber \\ 
& \quad + \int p(\mathbf{y}_R \vert \mathbf{g}_R, \mathbf{X}, \mathbf{H}_{\mathrm{BI}}, \mathbf{E})  \log \; p(\mathbf{y}_R \vert \mathbf{g}_R, \mathbf{X}, \mathbf{H}_{\mathrm{BI}}, \mathbf{E}) , \nonumber \\
&= \log \big[ \det \big( ( \mathbf{I}_N \otimes \mathbf{X}^H ) \mathbf{A} \mathbf{R}_{RC}  \mathbf{A}^H  ( \mathbf{I}_N \otimes \mathbf{X} )  +  \sigma_{R}^2 \mathbf{I}_{LN} \big) \big]  \nonumber \\ 
&\quad -  \log \big[ \det \big( ( \mathbf{I}_N \otimes \mathbf{X}^H ) \mathbf{A} \mathbf{R}_{C}  \mathbf{A}^H  ( \mathbf{I}_N \otimes \mathbf{X} )  +  \sigma_{R}^2 \mathbf{I}_{LN} \big) \big]  ,   \label{1-1} \\
&\approx \log \big[ \det \big( \frac{L}{\sigma_{R}^2} \widetilde{\mathbf{W}} \mathbf{A} \mathbf{R}_{RC}  \mathbf{A}^H  \widetilde{\mathbf{W}}^H + \mathbf{I}_{N(K+N)} \big) \big] 
\nonumber \\
& \quad - \log \big[ \det \big( \frac{L}{\sigma_{R}^2} \widetilde{\mathbf{W}} \mathbf{A} \mathbf{R}_{C}  \mathbf{A}^H  \widetilde{\mathbf{W}}^H + \mathbf{I}_{N(K+N)} \big) \big], \label{1-2}
\end{align}
where $\mathcal{H}(\cdot)$ denotes the differential entropy, $p(\cdot)$ represents the probability distribution function (PDF),  $ \mathbf{A} = \mathbf{H}_{\mathrm{BI}}^T \mathbf{E}^T \otimes \mathbf{H}_{\mathrm{BI}}^H \mathbf{E}^H$, $\widetilde{\mathbf{W}} = \mathbf{I}_N \otimes \mathbf{W}^H$, $\mathbf{R}_{RC} = \mathbf{R}_R + \mathbf{R}_C$, \eqref{1-2} follows the properties of determinant $\det(\mathbf{I}_a + \mathbf{A} \mathbf{B}) = \det(\mathbf{I}_b + \mathbf{B} \mathbf{A})$, $\det(c\mathbf{A}) = c^a \mathbf{A}$ \cite{zxd},
and the condition \eqref{approx} with sufficiently large $L$.

\begin{Remark}\normalfont
Sensing MI \eqref{1-2} is similar to the capacity in a non-regenerative MIMO relay system without direct link \cite{relayconference,relaytwc}. However, due to consideration of echo signal, \eqref{1-2} is more complicated,
containing quartic terms in reflection coefficients matrix
$\mathbf{E}$
and quadratic terms in BS beamformer
$\mathbf{W}$.
\end{Remark}

In the following,  we discuss two cases separately: one is the simplified case with the LoS BS-IRS channel and no clutter interference to sensing, while  the other one is the generalized case with the Rician BS-IRS channel  and the presence of clutter interference to sensing.

\vspace{-3mm}
\section{The simplified case: LoS BS-IRS Channel and no clutter interference to sensing}
In this section, assuming that the interfering echo signal is very weak, 
we can ignore the clutter interference, i.e., $\mathbf{R}_C = 0$.
Additionally,  if the BS and the IRS are deployed in the open space without significant  scatterers, 
we can assume that the channel between the BS and the IRS, $\mathbf{H}_{\mathrm{BI}}$,  is 
the LoS channel \cite{dailos}. 
As such  $\mathbf{H}_{\mathrm{BI}}$ can be simplified as 
\begin{align}
	\mathbf{H}_{\mathrm{BI}} = \xi \mathbf{b} (\theta_\mathrm{I}) \mathbf{a}^H (\theta_\mathrm{B}),    \label{los}
\end{align}
where $\theta_\mathrm{B}$ is the AOD relative to the BS,  $\theta_\mathrm{I}$ is the AOA relative to the IRS, $\mathbf{a} (\theta) \in \mathbb{C}^{N \times 1}$ is the transmit steering vector of the BS, and $\xi$ is the pathloss.
We aim to jointly design the BS beamforming matrix $\mathbf{W}$ and IRS reflection coefficients matrix $\mathbf{E}$ to maximize the sensing MI while satisfying users' individual  communication rate constraints, BS's transmit power constraint and unit modulus constraints of reflection elements. Thus, by letting  $\mathbf{R}_C = 0$ in \eqref{1-2},  the optimization problem is formulated as follows:
\begin{subequations} \label{3}
	\begin{align} 
		\max_{\mathbf{W}, \mathbf{E} } \quad  & \log \big[ \det \big( \frac{L}{\sigma_{R}^2} \widetilde{\mathbf{W}} \mathbf{A} \mathbf{R}_{R}  \mathbf{A}^H  \widetilde{\mathbf{W}}^H + \mathbf{I}_{N(K+N)} \big) \big] \label{3a} \\
		{\rm s.t.} \quad & R_k \geq r_k, \quad \forall k, \label{3b}\\
		&	{\Vert \mathbf{W} \Vert}_F^2 \leq P_0, \label{3c} \\
		&  	|e_m|^2 = 1, \quad m = 1,  \cdots, M, \label{3d}
	\end{align}
\end{subequations}
where $r_k$ is the required communication rate of the $k$-th user, and $P_0$ is the maximum BS transmit power.

Due to the high-degree polynomial of the determinant of the matrix and coupling variables, the nonconcave  objective function \eqref{3a} is difficult to tackle.
Hence, we first use the rank-one property of $\mathbf{H}_{\mathrm{BI}}$ 
 to transform the objective function \eqref{3a} into a tractable term. 

With \eqref{los}, matrix $\mathbf{A}$ can be rewritten as
\begin{align}
\mathbf{A} = (\mathbf{H}_{\mathrm{BI}}^T \otimes \mathbf{H}_{\mathrm{BI}}^H ) \mathbf{J}
= \xi^2 \overline{\mathbf{a}}(\theta_{\mathrm{B}})  \overline{\mathbf{b}}^H(\theta_{\mathrm{I}}) \mathbf{J}, \label{3-1}
\end{align}   
where $\overline{\mathbf{a}}(\theta_{\mathrm{B}}) = \big(\mathbf{a}^* (\theta_\mathrm{B}) \otimes \mathbf{a} (\theta_\mathrm{B}) \big) \in \mathbb{C}^{N^2 \times 1}$, $\overline{\mathbf{b}}^H(\theta_{\mathrm{I}}) = \big(\mathbf{b}^T (\theta_\mathrm{I}) \otimes \mathbf{b}^H (\theta_\mathrm{I})  \big) \in  \mathbb{C}^{1 \times M^2}$, $\mathbf{J} = \mathbf{E}^T \otimes \mathbf{E}^H$, and the derivation of \eqref{3-1} follows the property of the Kronecker product $  (\mathbf{A} \otimes \mathbf{C})(\mathbf{B} \otimes \mathbf{D}) = (\mathbf{AB}) \otimes (\mathbf{CD}) $ \cite{zxd}.

Then, substituting matrix $\mathbf{A}$ in \eqref{3a} with \eqref{3-1} and using the property of determinant $\det(\mathbf{I}_a + \mathbf{A} \mathbf{B}) = \det(\mathbf{I}_b + \mathbf{B} \mathbf{A})$ , the objective function \eqref{3a} can be transformed as 
\begin{align}
	 \log \big[ \det \big( \frac{L \xi^4}{\sigma_{R}^2}   \overline{\mathbf{b}}^H(\theta_{\mathrm{I}}) \mathbf{J} \mathbf{R}_{R} \mathbf{J}^H  \overline{\mathbf{b}}(\theta_{\mathrm{I}}) \overline{\mathbf{a}}^H(\theta_{\mathrm{B}}) 
	\widetilde{\mathbf{W}}^H \widetilde{\mathbf{W}} \overline{\mathbf{a}}(\theta_{\mathrm{B}})  + 1 \big)   \big]. \label{3-2}
\end{align}

In \eqref{3-2}, $\frac{L \xi^4}{\sigma_{R}^2} $ merely acts as a coefficient.
Hence,  Problem \eqref{3} can be simplified as 
\begin{subequations} \label{3-4}
	\begin{align}
		\max_{\mathbf{W}, \mathbf{E} } \quad  &
\overline{\mathbf{b}}^H(\theta_{\mathrm{I}}) \mathbf{J} \mathbf{R}_{R} \mathbf{J}^H  \overline{\mathbf{b}}(\theta_{\mathrm{I}}) \overline{\mathbf{a}}^H(\theta_{\mathrm{B}}) \widetilde{\mathbf{W}}^H \widetilde{\mathbf{W}} \overline{\mathbf{a}}(\theta_{\mathrm{B}})  \label{3-4a} \\
		{\rm s.t.} \quad &\eqref{3b}, \eqref{3c}, \eqref{3d}. \label{3-4b}
	\end{align}
\end{subequations}

%

Next, in order to address the coupling variables, we resort to an alternating optimization (AO) algorithm to solve Problem \eqref{3-4} in the following.

\vspace{-3mm}
\subsection{BS Beamforming Matrix Optimization}
In this subsection, we optimize  the BS beamforming matrix $\mathbf{W}$ when $\mathbf{E}$ is fixed.
Note that,  given $\mathbf{E}$, the term $\overline{\mathbf{b}}^H(\theta_{\mathrm{I}}) ( \mathbf{E}^T \otimes \mathbf{E}^H ) \mathbf{R}_{R} (\mathbf{E}^* \otimes \mathbf{E} )  \overline{\mathbf{b}}(\theta_{\mathrm{I}})$  in \eqref{3-4a} is a scalar.
Thus, maximizing \eqref{3a} is equivalent to maximizing $\overline{\mathbf{a}}^H(\theta_{\mathrm{B}}) \widetilde{\mathbf{W}}^H \widetilde{\mathbf{W}} \overline{\mathbf{a}}(\theta_{\mathrm{B}}) $, and this equation can be further transformed as 
\begin{align}
	 & \quad \big( \mathbf{a}^T(\theta_{\mathrm{B}}) \mathbf{a}^*(\theta_{\mathrm{B}})\big) \otimes 
        \big(\mathbf{a}^H(\theta_{\mathrm{B}}) \mathbf{W} \mathbf{W}^H \mathbf{a}(\theta_{\mathrm{B}})  \big)  \nonumber \\
     &= N \mathbf{a}^H(\theta_{\mathrm{B}}) \mathbf{W} \mathbf{W}^H \mathbf{a}(\theta_{\mathrm{B}}), \label{3-5}
\end{align}
where 
the derivation of \eqref{3-5} follows the property of Kronecker product $ (\mathbf{AB}) \otimes (\mathbf{CD}) = (\mathbf{A} \otimes \mathbf{C})(\mathbf{B} \otimes \mathbf{D}) $ and the equality  $\mathbf{a}^T(\theta_{\mathrm{B}}) \mathbf{a}^*(\theta_{\mathrm{B}}) = N$.


Then, we let $\mathbf{C}_R = \mathbf{W}_R \mathbf{W}_R^H$ and  $\mathbf{W}_k = \mathbf{w}_k \mathbf{w}_k^H, k =1, \cdots, K$. 
We also  let $\mathbf{h}_k^H = \mathbf{h}_{\mathrm{BU},k}^H + \mathbf{h}_{\mathrm{IU},k}^H  \mathbf{E} \mathbf{H}_{\mathrm{BI}}$ and $\Omega_k = 2^{r_k} - 1, k =1, \cdots, K$. 
By using  SDR and dropping rank-1 constraints \cite{sdr}, the optimization subproblem of Problem \eqref{3-4} corresponding to the BS beamforming matrix $\mathbf{W}$  can be reformulated as an semidefinite programming (SDP) problem  given by 
\begin{subequations} \label{3-7}
	\begin{align}
		\max_{\mathbf{C}_{R}, \mathbf{W}_1, \cdots, \mathbf{W}_K } \quad  & \mathbf{a}^H(\theta_{\mathrm{B}}) \big( \sum_{k = 1}^{K} \mathbf{W}_k + \mathbf{C}_R\big)  \mathbf{a}(\theta_{\mathrm{B}}) \label{3-7a} \\
		{\rm s.t.} \quad & \Omega_k \text{Tr}\big( \mathbf{h}_k \mathbf{h}_k^H  ( \sum_{k = 1}^{K} \mathbf{W}_k + \mathbf{C}_R) \big)  + \Omega_k \sigma_{k}^2   \nonumber \\
		& - (1+\Omega_k) \text{Tr}( \mathbf{h}_k \mathbf{h}_k^H  \mathbf{W}_k ) \leq 0, \quad \forall k, \label{3-7b} \\
		& \text{Tr}\big( \sum_{k = 1}^{K} \mathbf{W}_k + \mathbf{C}_R\big) \leq P_0, \label{3-7c} \\
		& \mathbf{C}_{R} \succeq 0, \quad \mathbf{W}_k \succeq 0, \quad \forall k. \label{3-7d}
	\end{align}
\end{subequations}
The problem above is a standard SDP problem which can be solved using CVX.

Let the optimal solution to Problem \eqref{3-7} be $\tilde{\mathbf{W}}_k$ and $\tilde{\mathbf{W}}_R$. The following lemma proves that Problem \eqref{3-7} always has the optimal solution of $\tilde{\mathbf{W}}_k, \forall k$ satisfying rank-1 constraint and $\tilde{\mathbf{W}}_R = \mathbf{0}$, i.e., 
the dedicated sensing beamformer is unnecessary for improving MI.

\begin{lemma} \label{lemma-1}
Problem \eqref{3-7} always has a rank-1 optimal solution $\tilde{\mathbf{W}}_k, \forall k$ and  sensing beamformer $\tilde{\mathbf{W}}_R = \mathbf{0}$.
\end{lemma}
\textbf{\textit{Proof: }} See Appendix \ref{proof1}. \hspace{4.8cm} $\blacksquare$

\vspace{-4mm}
\subsection{IRS Reflection Coefficients Optimization}
In this subsection, given $\mathbf{W}$, the optimization subproblem of Problem \eqref{3-4} corresponding to IRS reflection coefficients matrix $\mathbf{E}$ can be reformulated as 
\begin{subequations} \label{4-1}
	\begin{align}
	\min_{ \mathbf{E} } \quad  & - \overline{\mathbf{b}}^H(\theta_{\mathrm{I}}) ( \mathbf{E}^T \otimes \mathbf{E}^H ) \mathbf{R}_{R} (\mathbf{E}^* \otimes \mathbf{E} )  \overline{\mathbf{b}}(\theta_{\mathrm{I}})    \label{4-1a} \\
	{\rm s.t.} \quad &\eqref{3b},  \eqref{3d}. \label{4-1b}
	\end{align}
\end{subequations}

Then, denoting  $\mathbf{e}^H = [ e_1, e_2, \cdots, e_M] \in \mathbb{C}^{1 \times M}$ as the equivalent reflection coefficients vector, the communication  rate constraints \eqref{3b} can be rewritten as 
\begin{align}
	& 2 \text{Re}\big( \mathbf{e}^H 
	( \Omega_k \sum_{i=1, i\neq k }^{K} \mathbf{d}_{k,i}  - \mathbf{d}_{k,k} + \Omega_k \mathbf{d}_{k,r} )  \big) + l_k \nonumber \\
	&+ \mathbf{e}^H ( \Omega_k \sum_{i=1, i\neq k }^{K} \mathbf{D}_{k,i}  -  \mathbf{D}_{k,k} + \Omega_k \mathbf{D}_{k,r}) \mathbf{e} \leq 0, \quad  \forall k, \label{4-2}
\end{align}
where $l_k = \Omega_k \sum_{i=1, i\neq k }^{K}  c_{k,i} + \Omega_k \sigma_{k}^2 - c_{k,k} + \Omega_k c_{k,r}$, $c_{k,i} =$  \\ $ \mathbf{h}_{\mathrm{BU},k}^H \mathbf{w}_i \mathbf{w}_i^H \mathbf{h}_{\mathrm{BU},k}$,  $c_{k,r} = \mathbf{h}_{\mathrm{BU},k}^H \mathbf{W}_{\mathrm{R}} \mathbf{W}_{\mathrm{R}}^H \mathbf{h}_{\mathrm{BU},k}$,
$\mathbf{d}_{k,i} =$   \\ $  \mathrm{diag} (\mathbf{h}_{\mathrm{IU},k}^H) \mathbf{H}_{\mathrm{BI}} \mathbf{w}_i \mathbf{w}_i^H \mathbf{h}_{\mathrm{BU},k} $,
$\mathbf{d}_{k,r} =  \mathrm{diag} (\mathbf{h}_{\mathrm{IU},k}^H) \mathbf{H}_{\mathrm{BI}} \mathbf{W}_{\mathrm{R}} $  \\$ \mathbf{W}_{\mathrm{R}}^H  \mathbf{h}_{\mathrm{BU},k}$,
$\mathbf{D}_{k,i}  = \mathrm{diag} (\mathbf{h}_{\mathrm{IU},k}^H) \mathbf{H}_{\mathrm{BI}} \mathbf{w}_i \mathbf{w}_i^H \mathbf{H}_{\mathrm{BI}}^H  \mathrm{diag} (\mathbf{h}_{\mathrm{IU},k}^H)^H$,  and 
$\mathbf{D}_{k,r} = \mathrm{diag} (\mathbf{h}_{\mathrm{IU},k}^H) \mathbf{H}_{\mathrm{BI}}\mathbf{W}_{\mathrm{R}} \mathbf{W}_{\mathrm{R}}^H \mathbf{H}_{\mathrm{BI}}^H  \mathrm{diag} (\mathbf{h}_{\mathrm{IU},k}^H)^H$.


In order to transform the nonconvex constraints \eqref{4-2} into convex constraints,
EVD and the equation $\mathbf{e}^H \mathbf{e} = M$ are applied,  and then \eqref{4-2} can be further expressed as 
\begin{align}
\mathbf{e}^H \widetilde{\mathbf{F}_k} \mathbf{e} + \lambda_{min} ( \mathbf{F}_k ) M 
+ 2 \text{Re}( \mathbf{e}^H  \overline{\mathbf{d}}_k ) + l_k \leq 0, \quad \forall k, 
\label{4-3}
\end{align}
where $\overline{\mathbf{d}}_k =  \Omega_k \sum_{i=1, i\neq k }^{K} \mathbf{d}_{k,i}  - \mathbf{d}_{k,k} + \Omega_k \mathbf{d}_{k,r}$,
$\widetilde{\mathbf{F}_k} = \mathbf{F}_k - \lambda_{min} ( \mathbf{F}_k ) \mathbf{I}_M$, $\widetilde{\mathbf{F}_k} \succeq \mathbf{0}$  and $\mathbf{F}_k = \Omega_k \sum_{i=1, i\neq k }^{K} \mathbf{D}_{k,i}  -  \mathbf{D}_{k,k} + \Omega_k \mathbf{D}_{k,r} $.

By taking \eqref{R_R} into  \eqref{4-1a}, the objective function can be rewritten as
\begin{subequations}
\begin{align}
	& \quad   -\sum_{p=1}^{N_r}  \beta_p^2 \overline{\mathbf{b}}^H(\theta_{\mathrm{I}}) 
	\breve{\mathbf{E}} \breve{\mathbf{E}}^H \overline{\mathbf{b}}(\theta_{\mathrm{I}})    \nonumber \\
	&=   -\sum_{p=1}^{N_r}  \beta_p^2 \overline{\mathbf{b}}^H(\theta_{\mathrm{I}}) 
	\mathbf{C}_p ( \mathbf{e}^* \otimes \mathbf{e} ) ( \mathbf{e}^T \otimes \mathbf{e}^H  ) \mathbf{C}_p^H \overline{\mathbf{b}}(\theta_{\mathrm{I}}), \label{4-4a} \\
	&=    -( \mathbf{e}^T \otimes \mathbf{e}^H  )  \mathbf{M}_R  ( \mathbf{e}^* \otimes \mathbf{e} ), \label{4-4}
\end{align} 
\end{subequations}
where $\breve{\mathbf{E}} = ( \mathbf{E}^T \otimes \mathbf{E}^H )  \big(  \mathbf{b}^* (\theta_p^r)    \otimes \mathbf{b} (\theta_p^r) \big)$, $\mathbf{C}_p = \mathrm{diag}(\mathbf{b}^* (\theta_p^r)  ) $  \\ $ \otimes \mathrm{diag}(\mathbf{b} (\theta_p^r)  )$, $\mathbf{M}_R =  \sum_{p=1}^{N_r}  \beta_p^2 \mathbf{C}_p^H \overline{\mathbf{b}}(\theta_{\mathrm{I}}) \overline{\mathbf{b}}^H(\theta_{\mathrm{I}})   \mathbf{C}_p $,  
$\mathbf{M}_R \succeq \mathbf{0}$, \eqref{4-4a} follows  the property of Kronecker product $(\mathbf{A} \otimes \mathbf{B})(\mathbf{C} \otimes \mathbf{D})=(\mathbf{A}\mathbf{C}) \otimes (\mathbf{B}\mathbf{D})$ and the equation $\mathbf{E}^T \mathbf{b}^* (\theta)  = \mathrm{diag}(\mathbf{b}^* (\theta)  )  \mathbf{e}^* $.


It is challenging to deal with the nonconvex quartic objective function \eqref{4-4} with respect to (w.r.t.) the optimization  variable $\mathbf{e}$. Thus, we  adopt the MM method to deal with this problem \cite{mm}. 
More specifically, the MM method is aimed at constructing an easy-to-solve majorizer (minorizer)  for minimization (maximization) problem, where the majorizer must satisfy four conditions, i.e., $\mathrm{(A1)} - \mathrm{(A4)}$, given in \cite{multicast}.


Based on the MM framework and second-order Taylor expansion, the upper bound of the objective function \eqref{4-4} at the given point $\overline{\mathbf{e}}^{(\tau)}$ is given by
\begin{align}
\overline{\mathbf{e}}^H \mathbf{M}_1 \overline{\mathbf{e}} 
&\leq  \overline{\mathbf{e}}^H \text{Tr}(- \mathbf{M}_1)  \mathbf{I}_{M^2}  \overline{\mathbf{e}} + 2\text{Re} (  \overline{\mathbf{e}}^H \widetilde{\mathbf{M}}_1  \overline{\mathbf{e}}^{(\tau)}     ) + \mathrm{const}_1  \nonumber \\
	&= \text{Re} (  \overline{\mathbf{e}}^H \mathbf{g}_1 ) +  \mathrm{const}_2,  \label{4-6}
\end{align} 
where
$\mathbf{M}_1 = - \mathbf{M}_R, \mathbf{M}_1 \preceq \mathbf{0}$, $ \overline{\mathbf{e}} = \mathbf{e}^* \otimes \mathbf{e}$,
 $\widetilde{\mathbf{M}}_1 = \mathbf{M}_1 -  \text{Tr}(-\mathbf{M}_1) \mathbf{I}_{M^2} $,
$\mathrm{const}_1 = (\overline{\mathbf{e}}^{(\tau)})^H \widetilde{\mathbf{M}}_1     \overline{\mathbf{e}}^{(\tau)}$, $\mathrm{const}_2 = \text{Tr}(-\mathbf{M}_1)M^2 +  \mathrm{const}_1$, $\mathbf{g}_1 = 2 \widetilde{\mathbf{M}}_1    \overline{\mathbf{e}}^{(\tau)}$, the inequality follows from $\text{Tr}(-\mathbf{M}_1) \mathbf{I}_{M^2} \succeq \mathbf{M}_1$,
and the equality  follows from  $\overline{\mathbf{e}}^H \overline{\mathbf{e}} = M^2$.

For the purpose of unifying the optimization  variables, 
 we  express \eqref{4-6} as a function w.r.t. $\mathbf{e}$ rather than $\overline{\mathbf{e}}$.  Hence, the first term in \eqref{4-6} can be rewritten as 
\begin{align}
	\text{Re} (  \overline{\mathbf{e}}^H \mathbf{g}_1 ) = \text{Re} \big( ( \mathbf{e}^T  \otimes \mathbf{e}^H   ) \mathbf{g}_1 \big) = \text{Re} (  \mathbf{e}^H \mathbf{G}_1 \mathbf{e}  ),  \label{4-7}
\end{align}
where $\text{vec}(\mathbf{G}_1) = \mathbf{g}_1$, and the second equality follows the property of Kronecker product $(  \mathbf{C}^T \otimes \mathbf{A} ) \text{vec}(\mathbf{B})    = \text{vec}(\mathbf{A} \mathbf{B} \mathbf{C})$ \cite{zxd}.

Although matrix $\mathbf{G}_1 \in \mathbb{C}^{M \times M}$ can be recovered by rearranging the vector $\mathbf{g}_1 \in \mathbb{C}^{M^2 \times 1}$, the  computational cost for obtaining  $\mathbf{g}_1$ is much higher  compared to that for matrix $\mathbf{G}_1$. 
Thus, the following lemma shows the mathematical expression of $\mathbf{G}_1$.
\vspace{-3mm}
\begin{lemma} \label{lemma-2}
	Matrix $\mathbf{G}_1 \in \mathbb{C}^{M \times M} $ can be written as 
	\begin{align}
	\mathbf{G}_1 =	2 z_1 \mathbf{e}^{(\tau)} (\mathbf{e}^{(\tau)})^H  - 2  \sum_{p=1}^{N_r}  \beta_p^2 \mathbf{M}_3, \nonumber 
	\end{align} 
	where   $\mathbf{M}_3 = {\mathrm{diag}( \mathbf{b} (\theta_p^r)  ) }^H 
	\mathbf{b}(\theta_{\mathrm{I}})    \mathbf{b}^H(\theta_{\mathrm{I}})   \mathrm{diag}( \mathbf{b} (\theta_p^r)  )    \mathbf{e}^{(\tau)}  (\mathbf{e}^{(\tau)})^H $ \\ $  {\mathrm{diag}(\mathbf{b}(\theta_p^r)) }^H \mathbf{b}(\theta_{\mathrm{I}})  \mathbf{b}^H(\theta_{\mathrm{I}}) \mathrm{diag}( \mathbf{b} (\theta_p^r) )$ 
	and $z_1 =  -  \text{Tr}(-\mathbf{M}_1)$.
\end{lemma}
\textbf{\textit{Proof: }} See Appendix \ref{proof2}. \hspace{4.8cm} $\blacksquare$

Note that $\mathbf{G}_1$ is a Hermitian matrix, i.e., $\mathbf{e}^H \mathbf{G}_1 \mathbf{e} $ is a real number. However, $\mathbf{G}_1$ is a negative semidefinite matrix which makes \eqref{4-7} also nonconvex. In order to address this issue, we  resort to EVD, and  \eqref{4-7} can be transformed as
\begin{align}
 \mathbf{e}^H \mathbf{G}_1 \mathbf{e} =	\mathbf{e}^H \widetilde{\mathbf{G}_1} \mathbf{e} + \lambda_{min}(\mathbf{G}_1) M, \label{4-10}
\end{align}
where $\widetilde{\mathbf{G}_1} = \mathbf{G}_1 - \lambda_{min}(\mathbf{G}_1) \mathbf{I}_{M}$ and $\widetilde{\mathbf{G}_1} \succeq \mathbf{0}$.

Furthermore, by introducing nonnegative slack variable $\mathbf{f} = [f_1, f_2, \cdots, f_{2M}]^T$, we can adopt penalty CCP to tackle
nonconvex unit-modulus constraints \eqref{3d}, which are given by  \cite{ccp}
\begin{align}
	&|e_m|^2 \leq 1 + f_{M+m}, \quad m =1, \cdots, M, \label{4-11} \\
    & |e_m^{(\tau)}|^2 - 2\text{Re} (e_m^* e_m^{(\tau)}  ) \leq f_m -1 \quad m =1, \cdots, M. \label{4-12}
\end{align}

With \eqref{4-3}, \eqref{4-10}, \eqref{4-11} and \eqref{4-12},  Problem \eqref{4-1} can be recast as
\begin{subequations} \label{4-13}
	\begin{align}
		\min_{ \mathbf{e}, \mathbf{f} } \quad  &  \mathbf{e}^H \widetilde{\mathbf{G}_1} \mathbf{e} + \rho \sum_{i = 1}^{2M} f_i \label{4-13a} \\
		{\rm s.t.} \quad &\eqref{4-3},  \eqref{4-11}, \eqref{4-12}, \label{4-13b} \\
		& f_i \geq 0, \quad i =1, \cdots, 2M, \label{4-13c}
	\end{align}
\end{subequations}
where $\rho$ is the penalty parameter. Problem \eqref{4-13} is a convex second-order
cone programming (SOCP) problem which can be solved using CVX directly.

Finally, by letting  $g(\mathbf{W}, \mathbf{E})$ be the objective function value \eqref{3a}, the overall algorithm for solving Problem \eqref{3} is summarized in Algorithm \ref{alg1}.
\begin{algorithm}[htbp]
	\caption{Low-complexity Iterative Algorithm  for Solving Problem \eqref{3} }
	\label{alg1}
	\begin{algorithmic}[1]
		\STATE Set the outer iteration index $\kappa = 0 $. Initialize 
		$\mathbf{E}^{(0)}$ randomly, the tolerance $\epsilon_o$, and the maximum iterations number $\tau_{max}$.  
		\REPEAT
        \STATE Given $\mathbf{E}^{(\kappa)}$, solve Problem \eqref{3-7} and apply EVD to obtain BS beamforming matrix $\mathbf{W}^{(\kappa+1)}$.
	   	\STATE Set the inner iteration index $\tau = 0 $. Given $\mathbf{W}^{(\kappa+1)}$, obtain $\mathbf{E}^{(\kappa+1)}$ by the following steps.
		\REPEAT
		\IF {$\tau < \tau_{max}$} 
		\STATE Solve Problem \eqref{4-13} to obtain $\mathbf{e}^{(\tau+1)}$.
		\STATE Update $\rho = \min ( \mu\rho, \rho_{\mathrm{max}}  )$.
		\STATE Set $\tau = \tau+1$.
		\ELSE
		\STATE Return to step 4 and restart the iteration with a new initial random reflection coefficients matrix.
		\ENDIF
		\UNTIL $ | \frac{ g(\mathbf{W}^{(\kappa+1)}, \mathbf{E}^{(\tau+1)}) -  g(\mathbf{W}^{(\kappa+1)}, \mathbf{E}^{(\tau)}) }{g(\mathbf{W}^{(\kappa+1)}, \mathbf{E}^{(\tau)})} | \leq \epsilon_i$, ${\Vert \mathbf{e}^{(\tau+1)} - \mathbf{e}^{(\tau)} \Vert}_2 \leq \epsilon_d$ and ${\Vert \mathbf{f} \Vert}_1 \leq \epsilon_d  $.   
     	\STATE Let $\mathbf{E}^{(\kappa+1)} = \mathrm{diag}( (\mathbf{e}^{(\tau)})^* )$ and $ \kappa = \kappa + 1 $.
		\UNTIL $\left| \frac{ g(\mathbf{W}^{(\kappa+1)}, \mathbf{E}^{(\kappa+1)}) -  g(\mathbf{W}^{(\kappa)}, \mathbf{E}^{(\kappa)}) }{g(\mathbf{W}^{(\kappa)}, \mathbf{E}^{(\kappa)})} \right| \leq \epsilon_o$. 
	\end{algorithmic}
\end{algorithm} 

Note that the choice of $\rho$ may affect the convergence speed of Problem \eqref{4-13} \cite{ccp}. Thus, as shown in Step 8 of Algorithm  \ref{alg1},  we increase the value of $\rho$ by multiplying a ratio coefficient $\mu > 1$ in every iteration for the purpose of accelerating the convergence.
Furthermore, to avoid numerical issues, we also set an upper bound for $\rho$, i.e., $\rho < \rho_{\mathrm{max}}$. Additionally, in case that $\mathbf{E}^{(\kappa+1)}$ in Step 14 may not be feasible for Problem \eqref{4-1}, we can restart CCP procedure in Algorithm \ref{alg1} with a new random reflection coefficients matrix \cite{ccp}.

\subsection{Feasibility} \label{sec-fea}
In order to ensure the feasibility of Problem \eqref{3}, we need to firstly solve a power minimization problem, which is given by
      \begin{subequations} \label{fea-1}
    	\begin{align} 
    		\min_{\mathbf{W}, \mathbf{E} } \quad  & {\Vert \mathbf{W} \Vert}_F^2  \\
    		{\rm s.t.} \quad & R_k \geq r_k, \quad \forall k,  \\
    		&  	|e_m|^2 = 1, \quad m = 1, \cdots, M.
    	\end{align}
        \end{subequations}
The problem above is nonconvex and involves coupled variables, making it difficult to obtain a global optimal solution for it.  Hence, to get a good suboptimal solution, we can initialize a sufficiently large number of different sets of  random reflection coefficients, and then solve BS beamforming subproblems of Problem \eqref{fea-1} using SCA or SDR methods.
If the maximum power among these solutions is less than the maximum transmit power of the original problem, i.e., $P_0$, then  Problem \eqref{3} is feasible; otherwise, it is regarded as being infeasible.

\subsection{Complexity Analysis}
The computational complexity of Algorithm \ref{alg1} in every outer iteration is mainly due to  solving the SDP problem \eqref{3-7} and the penalty CCP algorithm from Steps 5 to 13.  The SDP problem is  solved by the interior point method in CVX, and the main computational complexity of the interior point method is given by  \cite{multicast}, \cite{complexitybook} \vspace{-2mm}
\begin{align}
	\mathcal{O} \Bigg(   \Big( \sum_{x = 1}^{X}  q_x + 2 Y \Big)^{\frac{1}{2}}  a   (  a^2 + a \sum_{x = 1}^{X}  q_x^2 + \sum_{x = 1}^{X}  q_x^3 + a \sum_{y = 1}^{Y} b_y^2) \Bigg), \nonumber
\end{align}
where $a$ is the number of variables, $X$ represents the number of linear matrix inequalities (LMIs) with dimension $q_x$, and $Y$ denotes the number of second-order cone (SOC) constraints with dimension $b_y$. Problem \eqref{3-7} is comprised of $2K+2$ LMIs with dimension $N^2$, and the number of variables is $a_1 = N(K+N)$. Hence,  the approximate complexity of solving Problem \eqref{3-7} is $\mathcal{O} \big(   [ (2K+2)N^2]^{\frac{1}{2}} a_1  [  a_1^2 +a_1(2K+2) N^4 + (2K+2)N^6  ]   \big)$. 
For the penalty CCP algorithm, the major complexity depends on calculating the maximum or minimum eigenvalue in \eqref{4-13a} and \eqref{4-3} and solving SOCP problem \eqref{4-13}. 
The complexity of the eigenvalue operation in the penalty CCP algorithm is $\mathcal{O} ( (K+1) M^3 )$. However, according to \cite{multicast} and \cite{complexitybook}, the complexity of solving SOCP problem is $\mathcal{O} ( UV^{3.5} + U^3 V^{2.5} )$, where $V$ represents the number of SOC constraints, and the dimension of each is $U$. In SOCP problem \eqref{4-13}, there are $K$ SOC constraints with dimension $M$ and $4M$ SOC constraints with dimension one. Therefore, the complexity of solving SOCP problem \eqref{4-13} per inner iteration is   $\mathcal{O} ( MK^{3.5} + M^3 K^{2.5} + (4M)^{3.5} +(4M)^{2.5} )$.
Finally, supposing that the iteration number of the CCP algorithm is $\tau_1$, the approximate computational complexity of Algorithm \ref{alg1} per outer iteration is $\mathcal{O} \big(   \tau_1 [ (K+1) M^3 + MK^{3.5} + M^3 K^{2.5} + (4M)^{3.5} +(4M)^{2.5}    ]  +  [ (2K+2)N^2]^{\frac{1}{2}} a_1  [  a_1^2 +a_1(2K+2) N^4 + (2K+2)N^6  ]    \big)$.

\subsection{Convergence Analysis}
Here, the convergence of Algorithm \ref{alg1} is analyzed. As mentioned in \cite{ccp},
when the conditions in Step 13 are satisfied, i.e., $ \sum_{i = 1}^{2M} f_i \approx 0 $, we have 
\begin{align}
 g(\mathbf{W}^{(\kappa)}, \mathbf{E}^{(\tau)}) \leq g(\mathbf{W}^{(\kappa+1)}, \mathbf{E}^{(\tau)}) \leq g(\mathbf{W}^{(\kappa+1)}, \mathbf{E}^{(\tau+1)}), \nonumber
\end{align}
where the first inequality follows  the optimality of SDP problem \eqref{3-7}, and the second inequality follows  the optimality of Problem \eqref{4-13}. Moreover, due to the bounded feasible set, the non-decreasing objective function \eqref{3a} has a finite upper bound. Therefore, the convergence of Algorithm \ref{alg1} is ensured.

\subsection{Sensing-Only System }
In this subsection, we consider the sensing-only system with the point target in the simplified case.
Next, we derive an upper bound on the sensing MI.
First, the sensing MI maximization problem can be formulated as 
$ \arg	\mathop{\max}\limits_{ \mathbf{W}, \mathbf{E}}  \quad \log \big[ \det \big( \frac{L}{\sigma_{R}^2} \widetilde{\mathbf{W}} \mathbf{A} \mathbf{R}_{R}  \mathbf{A}^H  \widetilde{\mathbf{W}}^H + \mathbf{I}_{N(K+N)} \big) \big]$, where  ${ \mathbf{W}, \mathbf{E} \in \{{\Vert \mathbf{W} \Vert}_F^2 \leq P_0, \quad
|e_m|^2 = 1, \forall m \} }$.
 Here,  a point target is considered, i.e., $ N_r = 1, \mathbf{R}_R
= \beta^2 (  \mathbf{b}^* (\theta^r)    \otimes \mathbf{b} (\theta^r) ) (  \mathbf{b}^* (\theta^r)    \otimes \mathbf{b} (\theta^r) )^H$. For simplicity, we omit the subscript $p$.
Then,  the following lemma provides an upper bound on sensing MI.
\begin{lemma} \label{lemma-3}
    In the sensing-only system with a point sensing target, the sensing MI can be upper-bounded by 
    \begin{align}
        &\quad \log \big[ \det \big( \frac{L}{\sigma_{R}^2} \widetilde{\mathbf{W}} \mathbf{A} \mathbf{R}_{R}  \mathbf{A}^H  \widetilde{\mathbf{W}}^H + \mathbf{I}_{N(K+N)} \big) \big] \nonumber \\ 
        &\leq 
        \log \big[ 
        \frac{L  \xi^4}{\sigma_{R}^2}  N^2 P_0 \beta^2 \tilde{\lambda}^2 M^2  + 1 \big], \label{lemma3-1} 
    \end{align}  
    where     $\tilde{\lambda} = \lambda_{max} (\mathbf{b} (\theta^r)  \mathbf{b}^H (\theta^r)   )$.  
    \end{lemma}
    \textbf{\textit{Proof: }} See Appendix \ref{proof3}. \hspace{4.8cm} $\blacksquare$

\section{The generalized case: Rician  BS-IRS Channel  and Presence of  clutter interference to sensing}
In this section, we consider a non-favorable propagation environment and 
 assume that the channel  between the BS and the IRS, $\mathbf{H}_{\mathrm{BI}}$, is a Rician channel, and the radar clutter interference cannot be neglected, i.e., \eqref{G_C} holds.
Hence, the optimization problem is formulated as 
\begin{subequations} \label{5}
	\begin{align} 
		\max_{\mathbf{W}, \mathbf{E} } \quad  & \log \big[ \det \big( \frac{L}{\sigma_{R}^2} \widetilde{\mathbf{W}} \mathbf{A} \mathbf{R}_{RC}  \mathbf{A}^H  \widetilde{\mathbf{W}}^H + \mathbf{I}_{N(K+N)} \big) \big] -  \nonumber \\ 
		&\log \big[ \det \big( \frac{L}{\sigma_{R}^2} \widetilde{\mathbf{W}} \mathbf{A} \mathbf{R}_{C}  \mathbf{A}^H  \widetilde{\mathbf{W}}^H + \mathbf{I}_{N(K+N)} \big) \big] \label{5a} \\
		{\rm s.t.} \quad & R_k \geq r_k, \quad \forall k, \label{5b}\\
		&	{\Vert \mathbf{W} \Vert}_F^2 \leq P_0, \label{5c} \\
		&  	|e_m|^2 = 1, \quad m = 1, \cdots, M. \label{5d}
	\end{align}
\end{subequations}

Note that Problem \eqref{5} is difficult to solve because of the nonconcave  objective function \eqref{5a}, nonconvex constraints \eqref{5b} and coupled variables, where 
the sensing MI \eqref{5a} with the consideration of radar clutter interference is more complicated than that given by  \eqref{3a} in Section \uppercase\expandafter{\romannumeral3}.
Furthermore, the determinant in \eqref{5a} is challenging to tackle due to high-rank  matrices $\mathbf{R}_{RC}$, $\mathbf{R}_C$ and $\mathbf{H}_{\mathrm{BI}}$, and the quartic term of $\mathbf{E}$ in \eqref{5a} is also difficult to cope with. As a result, we use the MM and SDR methods to transform the nonconvex problem \eqref{5} into a convex problem approximately \cite{mm}.

In order to use the MM method, we first make some transformations for the objective function \eqref{5a} denoted by $f(\mathbf{W}, \mathbf{E})$, which can be rewritten as  \vspace{-2mm}
\begin{subequations}
\begin{align}
	f(\mathbf{W}, \mathbf{E})  
	&=   -  \log \big[ \det {\big( \eta  (\mathbf{R}_{R}^{\frac{1}{2}})^H  \mathbf{P}^H \widetilde{\mathbf{T}}^{-1}  \mathbf{P} \mathbf{R}_{R}^{\frac{1}{2}}  + \mathbf{I}_{M^2} \big)}  ^{-1} \big], \label{5-2a} \\
	& = - \log \big[ \det \big(\mathbf{I}_{M^2} - \widetilde{\mathbf{P}}^H \mathbf{T}^{-1}  \widetilde{\mathbf{P}}  \big) \big], \label{5-2}
\end{align}
\end{subequations}
where $\eta = \frac{L}{\sigma_{R}^2}$, $\mathbf{P} = \widetilde{\mathbf{W}} \mathbf{A}$, $\widetilde{\mathbf{T}} = \mathbf{I}_{N(K+N)} + \eta \mathbf{P} \mathbf{R}_{C} \mathbf{P}^H  $, $\mathbf{T} = \mathbf{I}_{N(K+N)} + \eta \mathbf{P} \mathbf{R}_{RC} \mathbf{P}^H$, and $\widetilde{\mathbf{P}} = \sqrt{\eta} \mathbf{P} \mathbf{R}_{R}^{\frac{1}{2}}$. The derivation of \eqref{5-2a} follows the properties of determinant $\det(\mathbf{A}^{-1}) = \frac{1}{\det(\mathbf{A}) }$ and $\det(\mathbf{A}) \det(\mathbf{B}) = \det(\mathbf{A}  \mathbf{B})$, and \eqref{5-2} is derived from the Woodbury matrix identity \cite{horn}, i.e., $ (\mathbf{A}+\mathbf{U}\mathbf{C}\mathbf{V})^{-1} = \mathbf{A}^{-1} - \mathbf{A}^{-1}\mathbf{U}(\mathbf{C}^{-1}+\mathbf{V}\mathbf{A}^{-1}\mathbf{U})^{-1}\mathbf{V}\mathbf{A}^{-1} $.

Since $\mathbf{T}$ is a positive definite matrix, according to Lemma 3 in \cite{palomar}, the equation \eqref{5-2} denoted by $f(\widetilde{\mathbf{P}}, \mathbf{T})$ is jointly convex in $\{ \widetilde{\mathbf{P}}, \mathbf{T} \}$. 
Hence, similar to the derivation of the equation (32) in \cite{jin},
by using the first-order Taylor expansion, the minorizer  of $f(\widetilde{\mathbf{P}}, \mathbf{T})$ at the given point $( {\mathbf{P}}^{(n)} , {\mathbf{T}}^{(n)}   )$ is given by
\begin{align}
&  \eta 2\text{Re} \Big( \text{Tr} \big( \overline{\mathbf{A}}^{(n)}  \mathbf{P} \big) \Big)  
- \eta^2 \text{Tr} \big( \mathbf{B}^{(n)} \mathbf{P} \mathbf{R}_{RC} \mathbf{P}^H  \big)
 + \mathrm{const}_3, \label{5-3}
\end{align}
where $\overline{\mathbf{A}}^{(n)}  =    \mathbf{R}_{R}^{\frac{1}{2}}  (\mathbf{N}^{(n)})^{-1}  (\mathbf{R}_{R}^{\frac{1}{2}})^H  (\mathbf{P}^{(n)})^H (\mathbf{T}^{(n)})^{-1}$, 
$\mathbf{B}^{(n)} $ \\ $ =   (\mathbf{T}^{(n)})^{-1}    \mathbf{P}^{(n)}   \mathbf{R}_{R}^{\frac{1}{2}} (\mathbf{N}^{(n)})^{-1}  (\mathbf{R}_{R}^{\frac{1}{2}})^H  (\mathbf{P}^{(n)})^H  (\mathbf{T}^{(n)})^{-1}$,
$\mathbf{N}^{(n)} $ \\ $  =  \mathbf{I}_{M^2} - \eta  (\mathbf{R}_{R}^{\frac{1}{2}})^H   (\mathbf{P}^{(n)})^H (\mathbf{T}^{(n)})^{-1} \mathbf{P}^{(n)} \mathbf{R}_{R}^{\frac{1}{2}} $,
and
$\mathrm{const_3} =  \eta  \text{Tr}\big( \mathbf{B}^{(n)} \mathbf{T}^{(n)} \big)  -   \eta \text{Tr} \big( \mathbf{B}^{(n)}  \big) -  
 2 \eta \text{Re} \Big( \text{Tr} \big( \overline{\mathbf{A}}^{(n)}  \mathbf{P}^{(n)} \big) \Big) $.


After applying MM method,  \eqref{5-3} is also a nonconcave function of $(\mathbf{W}, \mathbf{E})$, especially the quartic function of $\mathbf{E}$. 
 Then, by utilizing some matrix transformations, we show that \eqref{5-3} is biconcave of $\text{vec}(\mathbf{W})$ and $\widetilde{\mathbf{E}}$ \cite{biconvex}, and  the SDR method can be  used, i.e., $\widetilde{\mathbf{E}} = \widetilde{\mathbf{e}}  \widetilde{\mathbf{e}}^H$ and $\widetilde{\mathbf{e}}^H = [ \mathbf{e}^H, 1 ] \in \mathbb{C}^{M+1}$. In other words, \eqref{5-3} is a concave function of $\text{vec}(\mathbf{W})$ with given $\widetilde{\mathbf{E}}$ and concave function of $\widetilde{\mathbf{E}}$ with given $\text{vec}(\mathbf{W})$. This motivates us to adopt an AO algorithm to update $\text{vec}(\mathbf{W})$ and $\widetilde{\mathbf{E}}$ alternately.

\vspace{-3mm}
\subsection{BS Beamforming Matrix Optimization}
Given $\widetilde{\mathbf{E}}$, 
we apply the MM approach to solve the BS beamforming matrix optimization subproblem. In the following, $\mathbf{W}^{(n,l)}$ denotes the fixed point of Taylor expansion in every iteration, where the subscript $n$ represents the $n$-th outer iteration of AO algorithm, and the subscript $l$ indicates the $l$-th inner iteration of the MM method.

First, the first three terms in \eqref{5-3} can be respectively rewritten  as 
\begin{align}
	&\text{Tr} \big( \overline{\mathbf{A}}^{(n,l)}  \mathbf{P} \big) = \text{vec}^H \big( (\overline{\mathbf{A}}^{(n,l)})^H  (\mathbf{A}^{(n,l)})^H  \big) \text{vec} \big( \widetilde{\mathbf{W}} \big), \label{5-4} \\
 &\text{Tr} \big( \mathbf{B}^{(n,l)} \mathbf{P} \mathbf{R}_{RC} \mathbf{P}^H  \big) = \text{vec}^H \big( \widetilde{\mathbf{W}} \big)  \breve{\mathbf{B}}^{(n,l)}
\text{vec} \big( \widetilde{\mathbf{W}} \big).  \label{5-6}
\end{align}
where $ \breve{\mathbf{B}}^{(n,l)} = {\big(\mathbf{A}^{(n,l)}  \mathbf{R}_{RC} (\mathbf{A}^{(n,l)})^H  \big)}^T  \otimes \mathbf{B}^{(n,l)}  $.

Based on \eqref{5-4}-\eqref{5-6}, the equation \eqref{5-3} can be expressed as the function of $\text{vec} ( \widetilde{\mathbf{W}})$. However, in order to perform the optimization over $\text{vec} ( \mathbf{W} )$, we need  to transform \eqref{5-3} into a concave function of  $\text{vec} ( \mathbf{W} )$. The relationship between $\text{vec} ( \widetilde{\mathbf{W}})$ and  $\text{vec} ( \mathbf{W} )$ is given  by the following lemma. 
\vspace{-3mm}
\begin{lemma} \label{lemma-4}
 By letting  $\overline{\mathbf{w}} =\text{vec} ( \mathbf{W} ) $, we have $\text{vec} ( \widetilde{\mathbf{W}}) = \bm{\Phi} \overline{\mathbf{w}}^*$, where 
 \begin{equation}
		\bm{\Phi} =\left[   
		\mathbf{C}_{1} \otimes \mathbf{I}_{\widetilde{K}},   
		\mathbf{C}_{2} \otimes \mathbf{I}_{\widetilde{K}},     
		\dots , 
		\mathbf{C}_{NN} \otimes \mathbf{I}_{\widetilde{K}}  
		\right]^T \mathbf{K}_{N\widetilde{K}},  \nonumber 
	\end{equation}
  $\mathbf{K}_{N\widetilde{K}} \in  \mathbb{R}^{N\widetilde{K} \times N\widetilde{K}}$ 
  is the commutation matrix, $\widetilde{K} = K+N$, and $\mathbf{C}_{i} = \text{unvec}_{N,N} (\mathbf{c}_i) \in \mathbb{R}^{N \times N}, i = 1, \cdots, NN$ with $\mathbf{c}_i \in  \mathbb{R}^{N^2 \times 1}$ being the $i$-th column of the identity matrix $\mathbf{I}_{N^2}$.
\end{lemma}
\textbf{\textit{Proof: }} 
The proof is similar to \cite[Appendix B]{jin}, and thus  omitted here. \hspace{6.5cm} $\blacksquare$

With the aid of Lemma \ref{lemma-4} and equations \eqref{5-4}-\eqref{5-6}, the minorizer \eqref{5-3} can be rewritten as
\begin{align}
	&2\eta \text{Re} \big(  \text{vec}^T \big( (\overline{\mathbf{A}}^{(n,l)})^H  (\mathbf{A}^{(n,l)})^H  \big) \bm{\Phi}^*  \overline{\mathbf{w}} \big) + - \eta^2 \overline{\mathbf{w}}^H \widetilde{\mathbf{B}} \overline{\mathbf{w}} + \mathrm{const}_3, \label{5-7}
\end{align}
where $\widetilde{\mathbf{B}} =  \bm{\Phi}^T  \Big( \big(\mathbf{A}^{(n,l)}  \mathbf{R}_{RC} (\mathbf{A}^{(n,l)})^H  \big) \otimes (\mathbf{B}^{(n,l)})^T  \Big)   \bm{\Phi}^* \succeq 0$.

Moreover, for the purpose of unifying optimization  variables in the BS beamforming matrix optimization subproblem, constraints \eqref{5b} and \eqref{5c} can be respectively transformed as 
\begin{align}
	 	 \overline{\mathbf{w}}^H  \widetilde{\mathbf{h}}_k \overline{\mathbf{w}}  + \Omega_k \sigma_{k}^2  &\leq 0, \quad \forall k, \label{5-8} \\
	 \overline{\mathbf{w}}^H  \overline{\mathbf{w}} &\leq P_0, \label{5-9} 
\end{align}  
where $\widetilde{\mathbf{h}}_k = \Omega_k \sum_{i=1, i\neq k }^{K} \mathbf{\Gamma}_{k,i}  +  \Omega_k  \widetilde{\mathbf{Q}}_k   - \mathbf{\Gamma}_k  $, $\mathbf{\Gamma}_k = {( \mathbf{i}_k^T \otimes \mathbf{I}_N  )}^H \mathbf{h}_k   \mathbf{h}_k^H ( \mathbf{i}_k^T \otimes \mathbf{I}_N  )$, $\mathbf{\Gamma}_{k,i} = 
{( \mathbf{i}_i^T \otimes \mathbf{I}_N  )}^H \mathbf{h}_k   \mathbf{h}_k^H ( \mathbf{i}_i^T \otimes \mathbf{I}_N  )
$, $\widetilde{\mathbf{Q}}_k = ( \mathbf{Q}^* \mathbf{Q}^T \otimes \mathbf{h}_k   \mathbf{h}_k^H )$, $\mathbf{h}_k^H = \mathbf{h}_{\mathrm{BU},k}^H + \mathbf{h}_{\mathrm{IU},k}^H  \mathbf{E} \mathbf{H}_{\mathrm{BI}}$, $\mathbf{Q}^T = [ \mathbf{0}_{N \times K}, \mathbf{I}_N  ]$, and $\mathbf{i}_k$ is the $k$-th column of the identity matrix $\mathbf{I}_{K+N}$.


Note that communication  rate constraints \eqref{5-8} are nonconvex, and successive convex approximation (SCA) method is applied to transform  them into  convex constraints, which are given by
\begin{align}
	& -2  \text{Re} \big( (\overline{\mathbf{w}}^{(n,l)})^H  \mathbf{\Gamma}_k  \overline{\mathbf{w}}   \big) + (\overline{\mathbf{w}}^{(n,l)})^H  \mathbf{\Gamma}_k  \overline{\mathbf{w}}^{(n,l)}    \nonumber \\
&+ \Omega_k \Big( \sum_{i=1, i\neq k }^{K} \overline{\mathbf{w}}^H \mathbf{\Gamma}_{k,i} \overline{\mathbf{w}}   + \sigma_{k}^2 + \overline{\mathbf{w}}^H  \widetilde{\mathbf{Q}}_k   \overline{\mathbf{w}} \Big) \leq 0,  \enspace \forall k. \label{5-10} 
\end{align}

After  discarding the constant in \eqref{5-7}, dividing \eqref{5-7} by $\eta$, substituting \eqref{5b} with \eqref{5-10}, replacing \eqref{5c} by \eqref{5-9}, and transforming the maximization problem into minimization problem, the optimization subproblem  of Problem \eqref{5} corresponding to BS beamforming matrix can be reformulated as 
\begin{subequations} \label{5-11}
	\begin{align} 
		\min_{\overline{\mathbf{w}} } \quad  &  
	- 2 	\text{Re} \Big(\overline{\mathbf{w}}^H \bm{\Phi}^T \text{vec}^* \big( (\overline{\mathbf{A}}^{(n,l)})^H  (\mathbf{A}^{(n,l)})^H  \big)\Big)
		+ \eta \overline{\mathbf{w}}^H \widetilde{\mathbf{B}} \overline{\mathbf{w}}  
		\label{5-11a} \\
		{\rm s.t.} \quad &\eqref{5-9}, \eqref{5-10}. \label{5-11b}
	\end{align}
\end{subequations}

As a result, Problem \eqref{5-11} becomes  an SOCP problem which can be solved using CVX.

\vspace{-3mm}
\subsection{IRS Reflection Coefficients Optimization}
When  BS beamforming matrix $\mathbf{W}$ is fixed, we need to  transform Problem \eqref{5} into a convex problem of $\widetilde{\mathbf{E}}$. As aforementioned, SDR method is applied to deal with the quartic term of $\mathbf{E}$ in the minorizer \eqref{5-3}. In order to facilitate this transformation, the following equality is used \cite{horn},
\begin{align}
	\text{Tr} ( \mathrm{diag}(\mathbf{a})^H \mathbf{B} \mathrm{diag}(\mathbf{c}) \mathbf{D}^T  )  = \mathbf{a}^H ( \mathbf{B} \circ \mathbf{D}  ) \mathbf{c}. \label{5-12}
\end{align}

By using \eqref{5-12}, the other terms except the constant in the minorizer \eqref{5-3} can be respectively rewritten as 
\begin{align}
	&\text{Tr} \big( \overline{\mathbf{A}}^{(n)}  \mathbf{P} \big) 
	= \mathbf{1}_{N(K+N)}^H ( \mathbf{V}^{(n)} \circ  (\overline{\mathbf{A}}^{(n)})^T  ) \mathbf{j}, \label{5-13} \\
	&\text{Tr} \big( \mathbf{B}^{(n)} \mathbf{P} \mathbf{R}_{RC} \mathbf{P}^H  \big) 
	= \mathbf{j}^H \big( (\mathbf{V}^{(n)})^H \mathbf{B}^{(n)} \mathbf{V}^{(n)}  \circ \mathbf{R}_{RC}^T \big)  \mathbf{j}.  \label{5-15}
\end{align}
where $\mathbf{J} = \mathbf{E}^T \otimes \mathbf{E}^H$, $\mathbf{V}^{(n)} = \widetilde{\mathbf{W}}^{(n)} (\mathbf{H}_{\mathrm{BI}}^T \otimes \mathbf{H}_{\mathrm{BI}}^H ) $, 
and $\mathbf{j}$ is the column vector comprised of the diagonal elements of  $\mathbf{J}$, i.e., $\mathbf{j} = \text{vec} ( [\widetilde{\mathbf{E}}]_{1:M,1:M} )$. 

Hence, given $\mathbf{W}$,
maximizing \eqref{5-3} is equivalent to minimizing the following expression, 
\begin{align}
	\eta \mathbf{j}^H  \overline{\mathbf{V}}^{(n)} \mathbf{j} - 2\text{Re} \big( \mathbf{1}_{N(K+N)}^H ( \mathbf{V}^{(n)} \circ  (\overline{\mathbf{A}}^{(n)})^T  ) \mathbf{j} \big), \label{5-16}
\end{align}
where $\overline{\mathbf{V}}^{(n)} = \big( (\mathbf{V}^{(n)})^H \mathbf{B}^{(n)} \mathbf{V}^{(n)}  \circ \mathbf{R}_{RC}^T \big) \succeq 0 $. Here, similar to \eqref{5-11a}, we discard the constant in \eqref{5-3} and divide the remaining terms by a coefficient $\eta$.

Then, with the use of SDR, communication  rate constraints \eqref{5b} can be transformed into 
\begin{align}
	&\Omega_k \Big(  \sum_{i=1, i\neq k }^{K} \text{Tr} ( \widetilde{\mathbf{E}} \mathbf{H}_k \mathbf{w}_i \mathbf{w}_i^H \mathbf{H}_k^H ) +  \text{Tr} ( \widetilde{\mathbf{E}} \mathbf{H}_k \mathbf{W}_{\mathrm{R}} \mathbf{W}_{\mathrm{R}}^H \mathbf{H}_k^H ) \nonumber \\
	&+ \sigma_{k}^2 \Big) 
	- \text{Tr} ( \widetilde{\mathbf{E}} \mathbf{H}_k \mathbf{w}_k \mathbf{w}_k^H \mathbf{H}_k^H ) \leq 0, \quad \forall k, \label{5-17}
\end{align}
where $\mathbf{H}_k = [\mathbf{H}_{\mathrm{BI}}^H \mathrm{diag}(\mathbf{h}_{\mathrm{IU},k}^H)^H , \mathbf{h}_{\mathrm{BU},k} ]^H$ represents the equivalent channel from the BS to the $k$-th user.

Therefore, replacing \eqref{5b} with \eqref{5-17}, transforming maximization problem into minimization problem and dismissing the rank-one constraint, the optimization subproblem of Problem \eqref{5} corresponding to IRS reflection coefficients matrix $\mathbf{E}$ is recast as 
\begin{subequations} \label{5-18}
	\begin{align} 
		\min_{\widetilde{\mathbf{E}} } \quad & \eqref{5-16}  \label{5-18a} \\
		{\rm s.t.} \quad &\eqref{5-17}, \nonumber \\
		& [\mathrm{Diag} (\widetilde{\mathbf{E}}  )]_{m}  = 1, \quad m =1, \cdots, M+1, \label{5-18b} \\
		& \widetilde{\mathbf{E}}  \succeq 0.
	\end{align}
\end{subequations}

Problem \eqref{5-18} is a standard SDP problem which can be solved using CVX. Upon obtaining the optimal solution to Problem \eqref{5-18} without rank-1 constraint, Gaussian randomization method can be adopted to construct a suboptimal rank-1 solution \cite{sdr}.
More specifically, denoting $\hat{\mathbf{E}}_1$ as the rank-relaxed solution to Problem \eqref{5-18}, its EVD
is given by $\hat{\mathbf{E}}_1 = \mathbf{U}_1 \mathbf{\Sigma}_1 \mathbf{U}_1^H$, where $\mathbf{\Sigma}_1$ is the diagonal matrix with diagonal elements being the eigenvalues of $\hat{\mathbf{E}}_1$, and every column in $\mathbf{U}_1$ is an eigenvector of $\hat{\mathbf{E}}_1$.
Next, 1000 random candidate vectors are generated as $ \hat{\mathbf{e}}_i =\mathrm{exp} \Big( j \angle \big(  \frac{\mathbf{U}_1 \mathbf{\Sigma}_1^{\frac{1}{2}} \mathbf{q}_i  }{[\mathbf{U}_1 \mathbf{\Sigma}_1^{\frac{1}{2}} \mathbf{q}_i]_{M+1}} \big)  \Big) $, $i = 1, \cdots, 1000$, where $\mathbf{q}_i$ follows complex Gaussian distribution with zero mean and unit variance, i.e., $\mathbf{q}_i \sim \mathcal{CN} (0, \mathbf{I}_{M+1} )$. Based on this, the suboptimal rank-1 solution to Problem \eqref{5-18} is selected from $\widehat{\mathbf{E}} = [\mathrm{diag} ( \hat{\mathbf{e}}_i^*  )]_{1:M,1:M}$ satisfying all constraints in Problem \eqref{5} and maximizing the sensing MI \eqref{5a}. Furthermore, a sufficiently large  times of Gaussian randomization are necessary to ensure the non-decreasing objective function \eqref{5a}.

In the end, the overall algorithm to solve Problem \eqref{5} is outlined in Algorithm \ref{alg2}.

\begin{algorithm}[htbp]
	\caption{Generalized Iterative Algorithm for Solving Problem \eqref{5} }
	\label{alg2}
	\begin{algorithmic}[1]
		\STATE Set the outer iteration index $n=0$. Initialize  $\mathbf{E}^{(0)}$, $\mathbf{W}^{(0)}$, and the tolerance $\epsilon_2$.
		\REPEAT
		\STATE Given $\mathbf{E}^{(n)}$, let the inner iteration index $l = 0$ and $\mathbf{W}^{(n,l)} = \mathbf{W}^{(n)}$.
		\REPEAT
		\STATE Obtain $\mathbf{W}^{(n,l+1)}$ by solving Problem \eqref{5-11}, and then $l = l +1$.
		\UNTIL $   \left| \frac{ f(\mathbf{W}^{(n,l+1)}, \mathbf{E}^{(n)}) -  f(\mathbf{W}^{(n,l)}, \mathbf{E}^{(n)}) }{f(\mathbf{W}^{(n,l)}, \mathbf{E}^{(n)})} \right| \leq \epsilon_2 $. 
		\STATE Let $\mathbf{W}^{(n+1)} = \mathbf{W}^{(n,l)}$. 
		\STATE Given $\mathbf{W}^{(n+1)}$, obtain the IRS reflection coefficients matrix without rank-1 constraint by solving Problem \eqref{5-18}. Then, the rank-1 solution $\mathbf{E}^{(n+1)}$ can be constructed by the Gaussian randomization method.
		\STATE Set $n = n+1$.
		\UNTIL   $\left| \frac{ f(\mathbf{W}^{(n+1)}, \mathbf{E}^{(n+1)}) -  f(\mathbf{W}^{(n)}, \mathbf{E}^{(n)}) }{f(\mathbf{W}^{(n)}, \mathbf{E}^{(n)})} \right| \leq \epsilon_2$.
	\end{algorithmic}
\end{algorithm} 


\vspace{-4mm}
\subsection{Feasibility and Initialization}
The feasibility of Problem \eqref{5} can be ensured by considering  a corresponding  power minimization problem, similarly  as in  Section~\ref{sec-fea}, and the details are  thus omitted here.  

In Algorithm \ref{alg2}, the initial point, $(\mathbf{W}, \mathbf{E})$, which greatly affects the performance of the converged suboptimal solution, can be obtained by ensuring the QoS of the communication task, excluding the sensing task.
First, the IRS reflection coefficients can be optimized by maximizing the weighted channel gain \cite{irscom}, i.e., $ \arg	\mathop{\max}\limits_{\mathbf{E} \in \{ |e_m|^2 = 1, \quad \forall m \} } \quad \sum_{k = 1}^K \omega_k {\Vert (\mathbf{h}_{\mathrm{BU},k}^H + \mathbf{h}_{\mathrm{IU},k}^H  \mathbf{E} \mathbf{H}_{\mathrm{BI}})\Vert}^2$, where $\omega_k = \frac{1}{\Omega_k \sigma_{k}^2}$.
This optimization problem can be solved by SDR and Gaussian randomization methods \cite{irscom}, with the details  omitted here due to space limitation.
Second, given $\mathbf{E}$, we just need to find a feasible BS beamforming matrix $\mathbf{W}$ to Problem \eqref{5}, where the feasible solution can be obtained by resorting to  SDR method and Theorem 1 in \cite{liuxiang}.

%

\vspace{-3mm}
\subsection{Convergence Analysis}
In this subsection, we analyze the convergence of Algorithm \ref{alg2}. Let the objective function \eqref{5a} and the minorizer \eqref{5-3} be denoted by 
$f (\mathbf{W}, \mathbf{E}) $ and $g ( \mathbf{W}^{(n,l)} | \mathbf{W}^{(n)}, \mathbf{E}^{(n)}  )  $, respectively.  Hence, in the inner iteration from Steps 4 to 6, given $\mathbf{E}^{(n)}$, we have \vspace{-2mm}
\begin{align}
	&f (\mathbf{W}^{(n)}, \mathbf{E}^{(n)} )   \overset{(a)}{=}  g ( \mathbf{W}^{(n,l)} | \mathbf{W}^{(n)}, \mathbf{E}^{(n)}  ) \nonumber \\
	&\overset{(b)}{\leq} g ( \mathbf{W}^{(n,l+1)} | \mathbf{W}^{(n)}, \mathbf{E}^{(n)}  ) \overset{(c)}{\leq} f (\mathbf{W}^{(n+1)}, \mathbf{E}^{(n)} ), \label{7-1}
\end{align}
where $\mathbf{W}^{(n)} = \mathbf{W}^{(n,l)} $, $\mathbf{W}^{(n+1)} = \mathbf{W}^{(n,l+1)}$, $(a)$ follows  the MM condition $\mathrm{(A1)}$ in \cite{multicast}, $(b)$ is due to the optimality of maximization problem \eqref{5-11}, and $(c)$ follows the MM condition  $\mathrm{(A2)}$ in \cite{multicast}. Furthermore, 
the non-decreasing minorizer \eqref{5-3} has a finite upper bound because of the bounded feasible set.
Therefore, the objective function \eqref{5a} in the inner iteration can converge to a finite value.

Then,  given $\mathbf{W}^{(n+1)}$, similarly to  \eqref{7-1}, the following relationship holds:  \vspace{-2mm}
\begin{align}
	&f (\mathbf{E}^{(n)}, \mathbf{W}^{(n+1)} ) = g ( \mathbf{E}^{(n)} | \mathbf{E}^{(n)},     \mathbf{W}^{(n+1)}  ) \nonumber \\
	& \leq g ( \mathbf{E}^{(n+1)} | \mathbf{E}^{(n)},     \mathbf{W}^{(n+1)}  ) \leq f (  \mathbf{E}^{(n+1)}, \mathbf{W}^{(n+1)} ). \label{7-2}
\end{align}
Due  to the bounded feasible set, the non-decreasing objective function \eqref{5a} has a finite upper bound, which prove that  the convergence of the sequence $f (\mathbf{E}^{(n)}, \mathbf{W}^{(n)} )$ in Algorithm \ref{alg2}.

\vspace{-4mm}
\subsection{Complexity Analysis}
The major complexity of Algorithm \ref{alg2} in every outer iteration lies in solving SOCP problem \eqref{5-11} by MM framework and SDP problem \eqref{5-18}. Similar to the complexity analysis of Algorithm \ref{alg1}, the approximate complexity of Problem \eqref{5-11} is $\mathcal{O} (  N(K+N)(K+1)^{3.5} + N(K+N)(K+1)^{2.5}  )$, and that  of Problem  \eqref{5-18} is given by $\mathcal{O} ( [(K+1)(M+1)]^{\frac{1}{2}} M [M^2 + M(K+1)(M+1)^2  +(K+1)(M+1)^3 ]   ) $. Hence, 
 the total complexity of Algorithm \ref{alg2} per outer iteration is approximately given by $\mathcal{O} (  l_2[N(K+N)(K+1)^{3.5} + N(K+N)(K+1)^{2.5}]  + [(K+1)(M+1)]^{\frac{1}{2}} M [M^2 + M(K+1)(M+1)^2  +(K+1)(M+1)^3 ]   ) $, where $l_2$ denotes the inner iteration number from Steps 4 to 6.

\section{Numerical Results}
In this section, numerical results are provided to evaluate the performance of our proposed algorithms in the IRS-aided MIMO ISAC system. Both the BS and the IRS employ ULA with $N = 8$ antennas and $M = 32$ reflecting elements, respectively. 
The Cartesian coordinate system is adopted in our paper, 
where the BS is located at $(\mathrm{0\thinspace m}, \mathrm{0\thinspace m}  )$, 
the IRS is located at $(\mathrm{50\thinspace m}, \mathrm{10\thinspace m}  )$, 
and communication users are randomly distributed in the circle centered at $(\mathrm{60\thinspace m}, \mathrm{0\thinspace m}  )$ with the radius of $\mathrm{2.5\thinspace m}$. 
The channels from the BS to the $k$-th user, $\mathbf{h}_{\mathrm{BU},k}$, and from the IRS to the $k$-th user, $\mathbf{h}_{\mathrm{IU},k}$, are assumed to be Rician fading with Rician factor $0.5$.  
Meanwhile, the channel between the BS and the IRS, $\mathbf{H}_{\mathrm{BI}}$, is considered in two cases:  LoS only as  the product of the steering vector of transceivers and Rician fading with Rician factor $0.5$.
The transmit/receive steering vector of the BS or the IRS is given by 
$\mathbf{a}(\theta) = \mathbf{b}(\theta) = [ 1, e^{-i \frac{2\pi d_1 }{\lambda_1} \sin \theta }, \dots, e^{-i \frac{2\pi (N_1 - 1)d_1 }{\lambda_1} \sin \theta }  ]$, where $\lambda_1$ is the wavelength, $N_1$ is the number of antennas or reflecting elements, and $d_1 = \frac{\lambda_1}{2}$ is the antenna or reflecting element spacing.
Next, the large-scale path loss coefficient is $ -30 - 10\alpha\log_{10}(d)$ dB, in which $d$ is the link length in meters and the path loss exponents $\alpha$ for the BS-IRS link, the IRS-users link and the BS-users link are respectively set as $2.5, 2.5$, and $3.5$.
We consider two types of the sensing targets: the point target, i.e., $N_r = 1$, located at the angle $\theta^r = 0^{\circ}$ with the average strength $\beta^2 = -20$ dB and the extended target, i.e., $N_r = 3$, with directions $\theta_1^r = -30^{\circ}, \theta_2^r = 0^{\circ}$ and $\theta_3^r = 30^{\circ}$ and average strength $\beta_1^2 = \beta_2^2 = \beta_3^2 = -20$ dB \cite{liuxiang}.
Furthermore, the clutter interference is located at $\theta_1^c = -80^{\circ}, \theta_2^c = -50^{\circ}$, $\theta_3^c = -10^{\circ}$, $\theta_4^c = 10^{\circ}, \theta_5^c = 50^{\circ}$,  $\theta_6^c = 80^{\circ}$, i.e., $N_c = 6$, with the average strength $\gamma_q^2 = 0 \,\text{dB}, q = 1,\cdots,N_c$ \cite{tangconstant}, \cite{cgl}.
The number of communication users are set as $K = 3$ with the required communication rate threshold $r_k = 3 \,\text{bps/Hz}, k = 1, \cdots, K$.
In addition, the power budget of the BS is $P_0 = 40 \,\text{dBm}$, the noise power is $\sigma_{R}^2 = \sigma_{k}^2 = -80 \,\text{dBm}, k = 1, \cdots, K $, and  the length of communication time slots, which is also the number of radar fast-time snapshots, is set to $L = 64$.
 The initial penalty coefficient $\rho$ is set to 10, with a ratio coefficient $\mu = 1.15$, and the maximum penalty coefficient $\rho_{\mathrm{max}} = 10^{10}$.  As for the stopping criteria in Algorithms \ref{alg1} and \ref{alg2}, we set $\epsilon_o = 10^{-7}, \epsilon_i = \epsilon_2 = 10^{-5}$ and $\epsilon_d = 10^{-3}$.

Our proposed schemes in Sections \uppercase\expandafter{\romannumeral3} and \uppercase\expandafter{\romannumeral4} are denoted by Algorithm 1 and 
Algorithm 2, respectively.
To further evaluate the performance of our proposed algorithms, the following benchmarks are used for comparison.
1) BP: This is the scheme  in Algorithm 2 of \cite{benchmark-bp}, based on the beampattern (BP) metric.
2) SNR:  This is the scheme   in \cite{lr-sinr}, where the sensing SNR metric is applied to design IRS-aided MIMO ISAC beamforming scheme.
 3)MI:  This is the existing MI-based scheme proposed as Algorithm 3 in \cite{irs-mi}.  
4) CBO: This approach optimizes the communication beamforming matrix only (CBO) with  $\mathbf{W}_{\mathrm{R}} = 0$ in our proposed algorithms. 
5) SO: In this scheme, IRS reflection coefficients  and BS beamformer  are optimized separately, i.e., separate optimization (SO). We firstly obtain $\mathbf{E}$  
by maximizing the weighted sum of the norms of the IRS cascaded channels in the desired sensing directions, i.e., $\mathop{\max}\limits_{\mathbf{E}} \sum_{p = 1}^{N_r} {\Vert 
\mathbf{b} (\theta_p^r)^H \mathbf{E} \mathbf{H}_{\mathrm{BI}}\Vert}^2  $.  
Next, $\mathbf{W}$
is optimized by solving Problem \eqref{5-11}.
 6) UB: This is the upper bound of sensing MI according to  Lemma \ref{lemma-3}.
7) Random: In this scenario, IRS reflection coefficients
are generated randomly within the range of $ [0, 2\pi )$. Then, the BS beamforming matrix can be obtained by solving Problem \eqref{5-11}.

\vspace{-4mm}
\subsection{Convergence Performance and Computational Time}
\begin{figure}[htbp]
	\centering \subfigure[Achievable MI versus the number of outer iterations.]{\includegraphics[width=2.8in,height=2.0in]{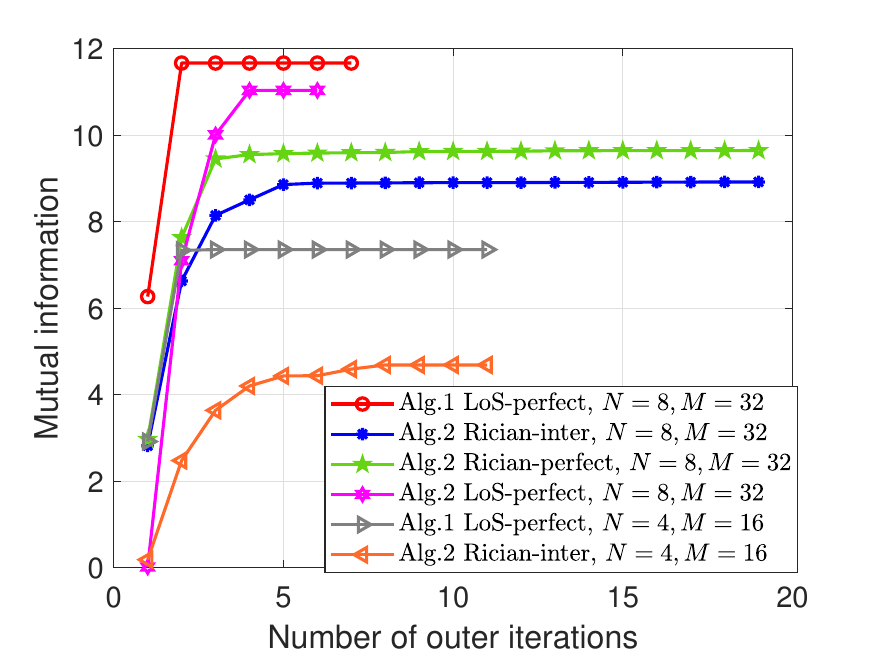}}
	\hspace{0.5cm} \subfigure[Achievable MI versus the number of inner iterations.]{\includegraphics[width=2.8in,height=2.0in]{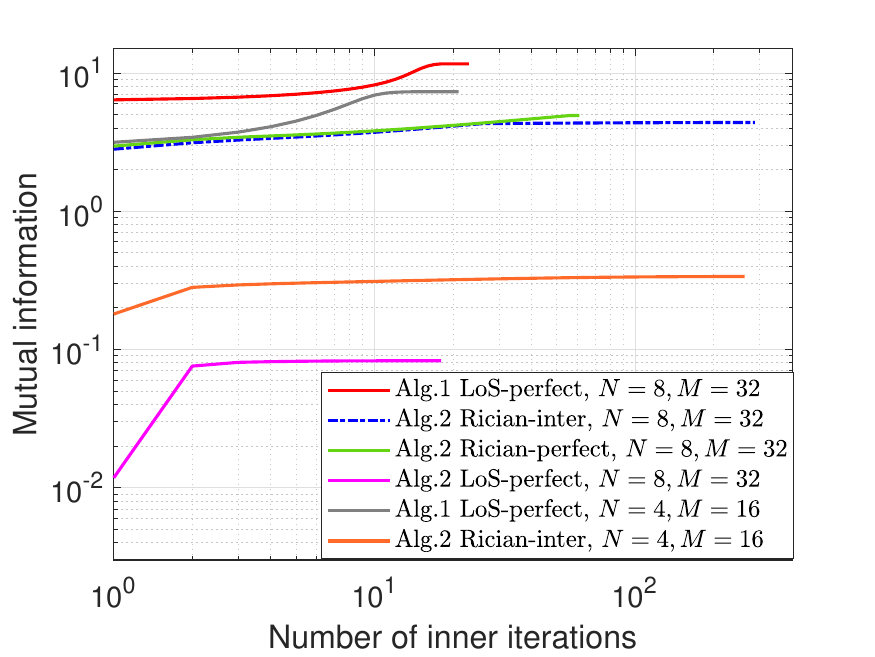}}
	\caption{Achievable MI versus the number of algorithm  iterations.}
	\label{iteration} 
\end{figure}

\begin{table}[ht]
    \centering 
    \begin{tabular}{|c|c|c|c|}
    \hline
    \textbf{Scenario} &    \textbf{CPU time (s)} \\
    \hline
    $\text{Alg.1 LoS-perfect}, N = 8, M = 32$   & 32.02\\
    $\text{Alg.2 Rician-inter}, N = 8, M = 32$   & 798.58\\
    $\text{Alg.2 Rician-perfect}, N = 8, M = 32$   & 824.66\\
    $\text{Alg.2 LoS-perfect}, N = 8, M = 32$   & 133.71\\
    $\text{Alg.1 LoS-perfect}, N = 4, M = 16$   & 18.85\\
    $\text{Alg.2 Rician-inter}, N = 4, M = 16$  & 157.02\\
    \hline
    \end{tabular}
    \caption{The computational time of the proposed algorithms required for convergence.}
    \label{table1}
\end{table}

In this subsection, we exhibit the convergence performance and the computational time of our proposed algorithms and consider the point sensing target.
When there is no clutter interference, i.e., $\mathbf{R}_C = 0$,
we denote Algorithm \ref{alg1} with $\mathbf{H}_{\mathrm{BI}}$ being LoS channel,  
Algorithm \ref{alg2} considering  LoS channel for $\mathbf{H}_{\mathrm{BI}}$ and  
Algorithm \ref{alg2} with  $\mathbf{H}_{\mathrm{BI}}$ being Rician channel
as "$\text{Alg.1 LoS-perfect}$", "$\text{Alg.2 LoS-perfect}$" and "$\text{Alg.2 Rician-perfect}$", respectively.  
Then, let "$\text{Alg.2 Rician-inter}$" represent the situation where  no clutter interference is present, and  $\mathbf{H}_{\mathrm{BI}}$ is Rician channel.

Fig. \ref{iteration} and table \ref{table1} illustrate the convergence performance and computational time of the proposed Algorithms \ref{alg1} and \ref{alg2}. As shown in \ref{iteration}, "Alg.1 LoS-perfect" converges to a better suboptimal point with five outer iterations and fewer inner iterations compared to "Alg.2 LoS-perfect", demonstrating its suitability in the simplified case.
For Rician channels, "Alg.2 Rician-perfect" achieves higher MI than "Alg.2 Rician-inter" with the same number of outer iterations and fewer inner iterations, indicating the impact of clutter interference on sensing performance. 
Additionally, the computational time of "Alg.1 LoS-perfect" is approximately $32\,\text{s}$, significantly faster than the $134\,\text{s}$ required by "Alg.2 LoS-perfect". Furthermore, increasing the number of BS antennas and IRS reflecting elements from $N=4, M=16$ to $N=8, M=32$ can enhance sensing MI but also lead to increased computational time.

\subsection{IRS-aided ISAC Beamforming Design in  Simplified Case}

In this subsection, 
we investigate the performance of the situation where  clutter interference is not present,  and LoS channel is considered  for $\mathbf{H}_{\mathrm{BI}}$.  Unless otherwise specified, the point sensing target is considered.

\begin{figure}[htbp]
	\centering \includegraphics[width=3in,height=2.2in]{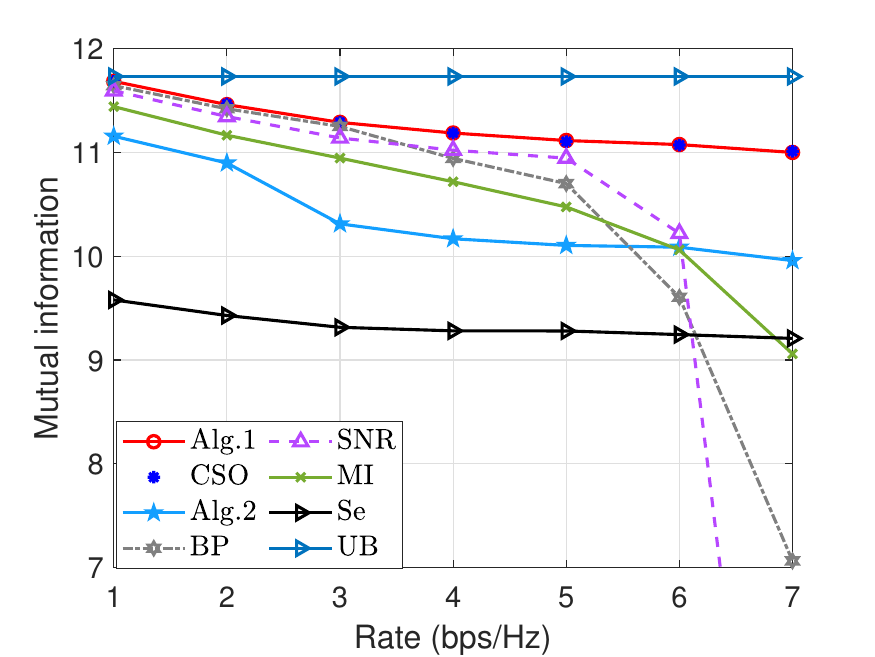}
	\caption{Achievable MI versus the communication rate requirement.}
	\label{rate-mi} 
\end{figure}


Fig. \ref{rate-mi} plots the achievable MI versus the  communication rate requirement. 
As proven in Lemma \ref{lemma-1}, the sensing MI of obtained by Algorithm \ref{alg1} is consistent with that of CBO, further demonstrating that in the simplified case, the dedicated sensing beamformer is unnecessary and  rank-1 communication beamformers always exist.
Compared to other schemes, Algorithm 1 can achieve higher sensing MI, especially when $r = 1 \, \text{bps/Hz}$, where it is almost identical to the upper bound of MI in the sensing only system. This demonstrates the superiority of Algorithm \ref{alg1}.
As the communication rate increases, the sensing MI of all schemes gradually decreases, except for the upper bound in the sensing only system. 
This is due to the tradeoff between the sensing and communication performance in the IRS-aided ISAC system, i.e.,  the increasing demand of communication can bring negative impacts on the sensing performance.
Then, when $r \geq 5 \, \text{bps/Hz}$, the decrease in Algorithms \ref{alg1} and \ref{alg2} is relatively mild, while the BP-based and SNR-based schemes decrease sharply. This indicates that the proposed schemes still perform well under high communication demands.

\begin{figure}[htbp]
	\centering \includegraphics[width=3in,height=2.2in]{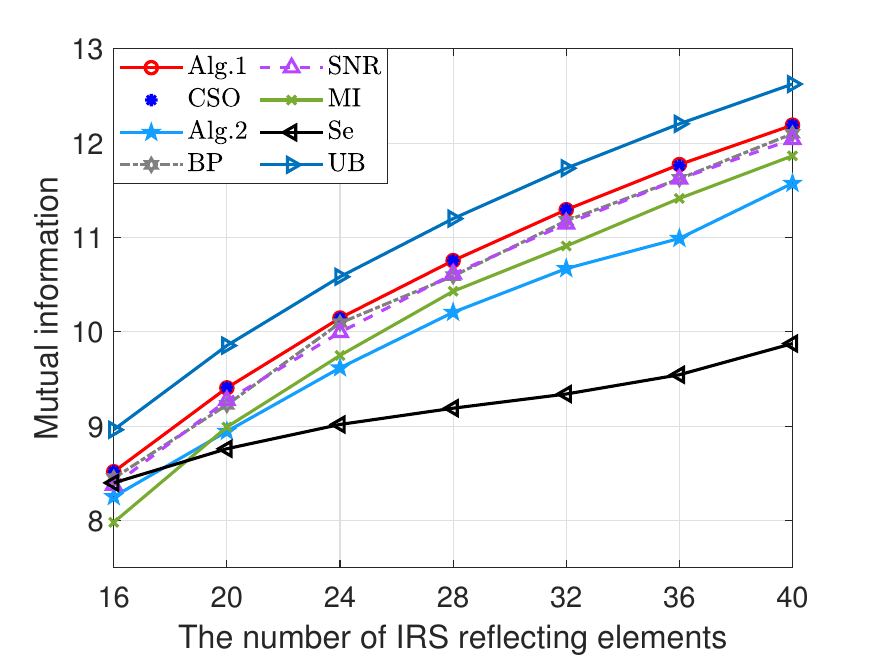}
	\caption{Achievable MI versus the number of IRS reflecting elements.}
	\label{RE-mi} 
\end{figure}

Fig. \ref{RE-mi} shows the achievable MI versus the number of IRS reflecting elements.
 First, MI of all schemes increases as the number of reflecting elements increases, which exhibits 
more MI gain from more IRS reflecting elements
and signifies  again the effects of IRS reflecting coefficients on the sensing performance. 
Secondly, Algorithm \ref{alg1} is exactly equal to the CBO and outperforms all other schemes, including Algorithm \ref{alg2}, further confirming Lemma \ref{lemma-1} and the superior performance of Algorithm \ref{alg1}. It can be seen that there is a noticeable gap between the mutual information of Algorithm \ref{alg1} and the upper bound, which is due to the impact of the communication task.

%
%

\begin{figure}[htbp]
	\centering \includegraphics[width=3in,height=2.2in]{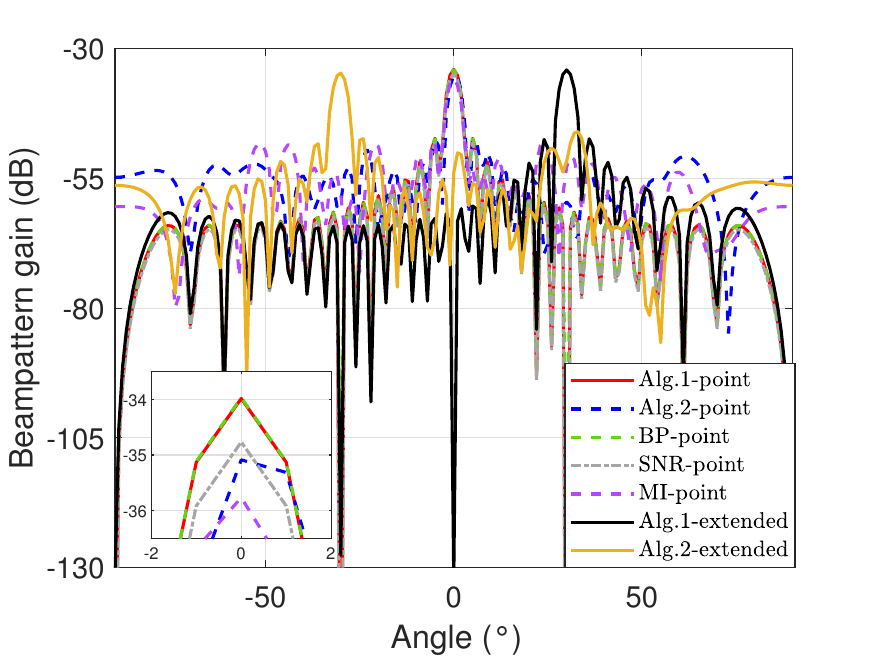}
	\caption{Beampatterns without clutter interference in the scenario of   LoS channel for $\mathbf{H}_{\mathrm{BI}}$.}
	\label{LOS-bp} 
\end{figure}

Fig. \ref{LOS-bp} illustrates the beampatterns of different schemes in the simplified case with no clutter interference and $\mathbf{H}_{\mathrm{BI}}$ being LoS channel.
"$\text{Alg.1-extended}$" and "$\text{Alg.2-extended}$" respectively represent the cases when the problem with  an extended target is solved by Algorithm \ref{alg1} and Algorithm \ref{alg2}.
In the scenario of the point target, the beampattern of "$\text{Alg.1-point}$" is almost the same as that of "$\text{BP-point}$" and has the highest mainlobe directed to the target angle $0^\circ$ with about $1 \, \text{dB}$ or $2 \, \text{dB}$ higher than the SNR-based, Algorithm \ref{alg2}-based, and MI-based schemes.
This demonstrates that the proposed Algorithm \ref{alg1} has the best sensing performance compared to other benchmarks.
However, from "$\text{Alg.1-extended}$" and "$\text{Alg.2-extended}$", we can observe that 
the extended target cannot be sensed thoroughly, 
and the beampatterns can only point to one direction with the other two sensing directions unperceived.
This is because the rank-1 LoS channel $\mathbf{H}_{\mathrm{BI}}$  can only distinguish one sensing parameter.

\subsection{IRS-aided ISAC Beamforming Design in Generalized Case}
In this subsection, we consider the extended sensing target and Rician channel for 
$\mathbf{H}_{\mathrm{BI}}$.

\begin{figure}[htbp]
	\centering \includegraphics[width=3in,height=2.2in]{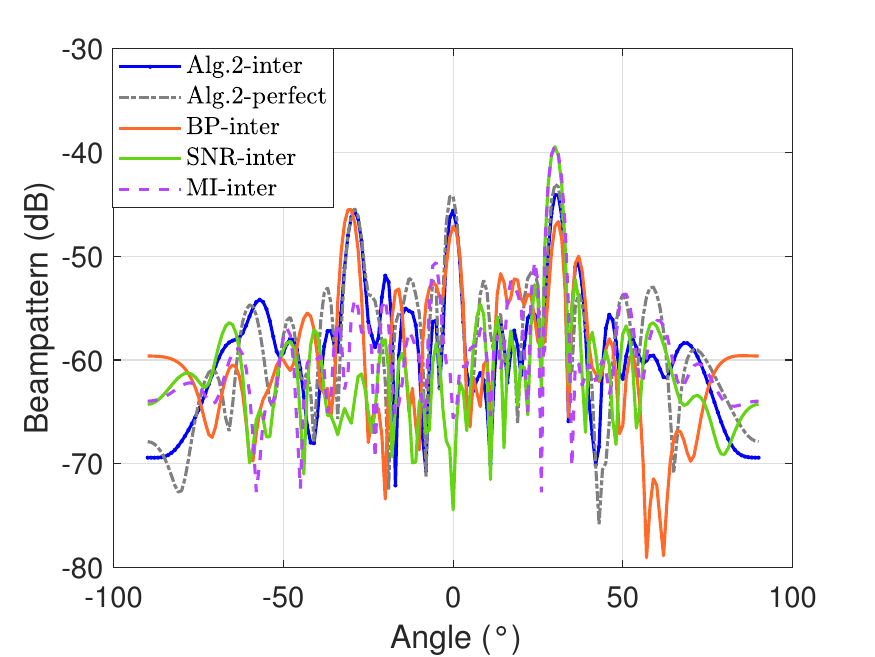}
	\caption{Beampatterns with clutter interference in the scenario of   
		Rician channel for $\mathbf{H}_{\mathrm{BI}}$.}
	\label{Rician-bp} 
\end{figure}


Fig. \ref{Rician-bp} compares the beampatterns of different algorithms.
The beamforming problem  without clutter interference solved by Algorithm \ref{alg2} is denoted as  "$\text{Alg.2-perfect}$", and the other schemes with clutter interference are respectively denoted as "$\text{Alg.2-inter}$", "$\text{BP-inter}$", "$\text{SNR-inter}$" , and "$\text{MI-inter}$".
First, compared to "$\text{Alg.1-extended}$" in Fig. \ref{LOS-bp}, the beampatterns of  "$\text{Alg.2-perfect}$" and "$\text{Alg.2-inter}$" can identify all  directions of the extended target, which shows 
that Rician channel with high rank can provide more gain for parameter sensing compared with the LoS channel.
Second, in presence of clutter interference, the beampattern of "$\text{SNR-inter}$" only has one mainlobe directed to the angle $30^{\circ}$ while the other desired sensing angles are missed.
By contrast, our proposed scheme based on Algorithm \ref{alg2} can perceive every sensing direction and has better beampattern than "$\text{SNR-inter}$", "$\text{BP-inter}$" and "$\text{MI-inter}$". 
Third, from the comparison between "$\text{Alg.2-inter}$" and "$\text{Alg.2-perfect}$", the mainlobe of "$\text{Alg.2-inter}$" is lower than that of "$\text{Alg.2-perfect}$", and the sidelobes of "$\text{Alg.2-inter}$" in some angles of clutter interference, e.g., $10^{\circ}, 50^{\circ}$ and $80^{\circ}$, are suppressed effectively. 
This illustrates that our proposed  Algorithm \ref{alg2} can achieve good beampattern while suppressing the clutter interference effectively.

\begin{figure}[htbp]
	\centering \includegraphics[width=3in,height=2.2in]{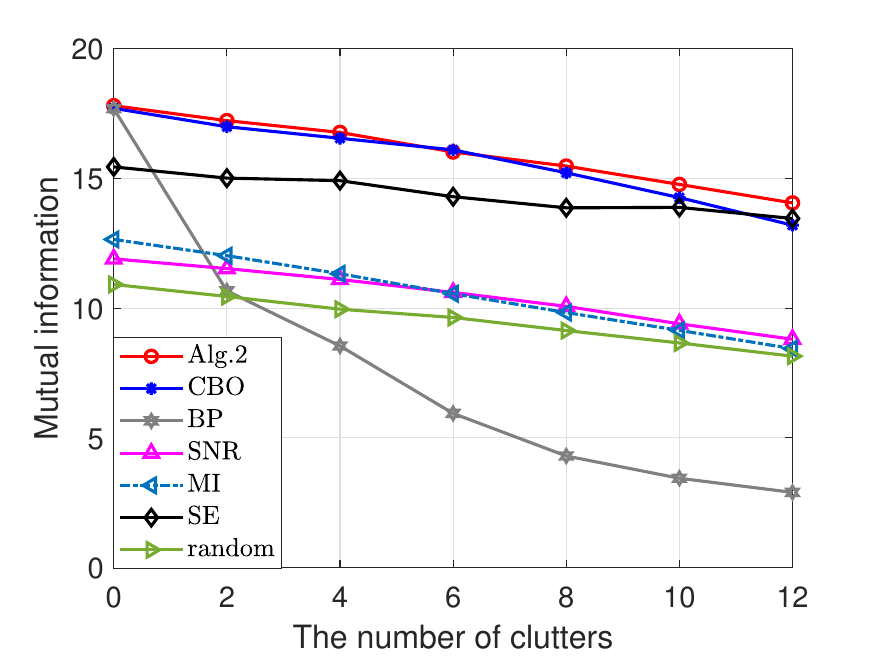}
	\caption{Achievable MI versus the number of clutters.}
	\label{clutter-mi} 
\end{figure}

 Fig. \ref{clutter-mi} depicts the achievable MI versus the number of clutters $N_c$.
All considered  schemes have the highest value of MI when  $N_c = 0$, i.e., there is no clutter interference, and their achievable MI  gradually degrades with the increase of the number of clutters, especially for BP which has a sharp decline of MI. 
This shows that more interfering scatterers can cause higher performance loss  on MI.
Furthermore, the MI of Algorithm \ref{alg2} is slightly higher than that of CBO, especially when the number of clutters exceeds 8, while significantly outperforming the other benchmarks. This demonstrates that a dedicated sensing beamformer can enhance MI in the presence of strong clutter interference, and our proposed Algorithm \ref{alg2} is more robust to clutter interference compared to other schemes.

\begin{figure}[htbp]
	\centering \includegraphics[width=3in,height=2.2in]{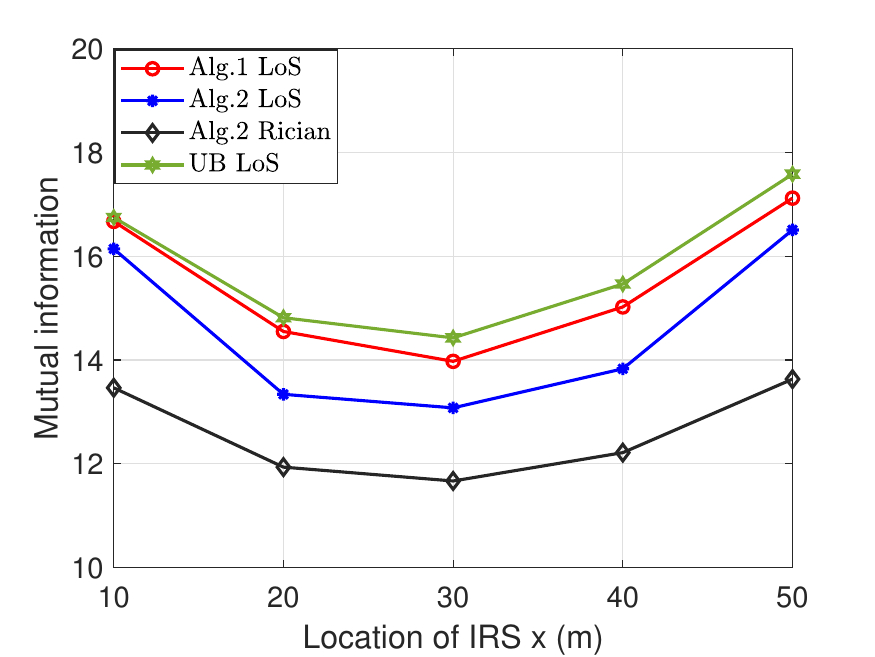}
	\caption{Achievable MI versus the location of the IRS $x$.}
	\label{locations-mi} 
\end{figure}

Fig. \ref{locations-mi} investigates the impact of IRS's location on the sensing MI.
The position of the IRS is set to $(\mathrm{x\thinspace m}, \mathrm{5\thinspace m}  )$, and 
we consider the sensing target as a point target located at $(\mathrm{50\thinspace m}, \mathrm{0\thinspace m}  )$ without clutter interference.
The radar range equation is given by $P_r = \frac{P_t G^2 \lambda^2 \sigma}{ 64 \pi^3 R^4 }$, where $P_r$ and $P_t$ are transmit and received power, $G$ is the antenna gain, $\lambda$ is the wavelength, $\sigma$ is the radar cross section, and $R$ is the distance between the BS and the scatterer \cite{swerling}. 
So, the average strength can be written as $ \beta^2 = \frac{G^2 \lambda^2 \sigma}{ 64 \pi^3 R^4 }$. The antenna gain, wavelength and RCS are set as  $G = 40 $ dB, $\lambda = 0.2 \, \text{m}$ and $\sigma = 1 \, \text{m}^2$, respectively \cite{swerling}.   
"$\text{Alg.1 LoS}$", "$\text{Alg.2 LoS}$", "$\text{Alg.2 Rician}$", and "$\text{UB LoS}$" respectively 
denote Algorithm \ref{alg1} with $\mathbf{H}_{\mathrm{BI}}$ being LoS channel,  
Algorithm \ref{alg2} considering  LoS channel for $\mathbf{H}_{\mathrm{BI}}$,
Algorithm \ref{alg2} with  $\mathbf{H}_{\mathrm{BI}}$ being Rician channel, 
and the upper bound in Lemma \ref{lemma-3}.
When the IRS is farther from the BS and sensing target, the sensing MI is low. However, when deployed near the BS or the target, the MI performs better. "$\text{Alg.1 LoS}$" outperforms other schemes and has a smaller gap to the upper bound, highlighting its superior performance. Additionally, "$\text{Alg.2 Rician}$" has lower sensing MI than "$\text{Alg.2 LoS}$", indicating that increased channel randomness negatively impacts the MI.

\vspace{-3mm}
\section{Conclusions}
In this paper, we studied an IRS-aided MIMO ISAC system based on sensing MI, by utilizing IRS as a colocated MIMO radar.
We considered  two cases:  the simplified case with 
LoS BS-IRS channel  and no clutter interference to sensing, and the generalized case with 
Rician fading channel for the BS-IRS link and the presence of clutter interference to sensing.
For both cases, we formulated the optimization problem by maximizing the sensing MI, subject to the QoS constraints for all communication users, the transmit  power constraint at the BS, and the unit-modulus constraint on the IRS's passive reflection. 
In the first case, we proved that the dedicated sensing beamformer is unnecessary, while the optimal rank-1 communication beamformer always exists, and we proposed a low-complexity iterative algorithm to solve the formulated optimization problem.
In the second case,
an alternative iterative algorithm, which can also be applied to the first case, 
was provided to solve the  problem under the  general setup. 
Numerical results were presented to demonstrate the performance superiority of proposed algorithms in comparison to various benchmark schemes.

\appendices{}    

\section{ THE PROOF OF LEMMA \ref{lemma-1} \label{proof1}   }
First, let the dual variables associated with $K$ SINR constraints \eqref{3-7b} and one power constraint \eqref{3-7c} denoted by $\mu_k, k = 1, 2,\cdots, K$ and $\mu_P$, respectively. We also associate matrix dual variables $\mathbf{L}_k, k = 1, 2,\cdots, K$ and $\mathbf{L}_R$ to $K+1$ positive semidefinite cone constraints \eqref{3-7d}.
Then, the Lagrangian function  and 
complimentary conditions of Problem \eqref{3-7} are given by
\begin{align}
    & \mathcal{L} = - \mathbf{a}^H(\theta_{\mathrm{B}}) \big( \sum_{k = 1}^{K} \mathbf{W}_k + \mathbf{C}_R\big)  \mathbf{a}(\theta_{\mathrm{B}})  +  \mu_k \big( \Omega_k \text{Tr}( \overline{\mathbf{H}}_k   \sum_{k = 1}^{K} \mathbf{W}_k )  \nonumber \\ 
    &+  \Omega_k \text{Tr}( \overline{\mathbf{H}}_k \mathbf{C}_R )  + \Omega_k \sigma_{k}^2 
    - (1+\Omega_k) \text{Tr}(  \overline{\mathbf{H}}_k  \mathbf{W}_k )  \big)  \nonumber \\ 
    &+
    \mu_P \big( \text{Tr}( \sum_{k = 1}^{K} \mathbf{W}_k + \mathbf{C}_R ) - P_0\big) - \sum_{k = 1}^{K} \text{Tr} ( \mathbf{L}_k \mathbf{W}_k ) \nonumber \\ 
    & - \text{Tr} ( \mathbf{L}_R \mathbf{C}_{R}),  \label{lemma1-1a} \\
    &  \mu_k \big( \Omega_k \text{Tr}( \overline{\mathbf{H}}_k   \sum_{k = 1}^{K} \mathbf{W}_k )  +  \Omega_k \text{Tr}( \overline{\mathbf{H}}_k \mathbf{C}_R )  + \Omega_k \sigma_{k}^2    \nonumber \\
		& - (1+\Omega_k) \text{Tr}(  \overline{\mathbf{H}}_k  \mathbf{W}_k )  \big) = 0, \quad \forall k, \label{lemma1-1b} \\
		& \mu_P \big( \text{Tr}( \sum_{k = 1}^{K} \mathbf{W}_k + \mathbf{C}_R ) - P_0\big) = 0, \label{lemma1-1c} \\
		& \text{Tr} ( \mathbf{L}_R \mathbf{C}_{R}) =  0, \quad \text{Tr} ( \mathbf{L}_k \mathbf{W}_k ) = 0, \quad \forall k. \label{lemma1-1d} \\
        & \mu_k \geq 0, \mathbf{L}_k \succeq 0, \, \forall k, \quad \mu_P \geq 0, \mathbf{L}_R \succeq 0, \label{lemma1-1e} \\ 
        & \eqref{3-7b}, \eqref{3-7c}, \eqref{3-7d}, \label{ori-pro}
\end{align}
where $\overline{\mathbf{H}}_k = \mathbf{h}_k \mathbf{h}_k^H $.

According to KKT conditions, the derivative of the Lagrangian function at the optimal point is zero, i.e., $\frac{\partial \mathcal{L}}{\partial \mathbf{W}_k} = 0, \forall k$ and $\frac{\partial \mathcal{L}}{\partial \mathbf{C}_R} = 0$. Thus, we can obtain
\begin{align}
   & \mathbf{L}_k = - \tilde{\mathbf{A}}(\theta_{\mathrm{B}}) - (\mu_k + \mu_k \Omega_k) \overline{\mathbf{H}}_k + \sum_{i = 1}^{K} \mu_i \Omega_i \overline{\mathbf{H}}_i + \mu_P \mathbf{I}_N,  \label{lemma1-2} \\
   & \mathbf{L}_R = - \tilde{\mathbf{A}}(\theta_{\mathrm{B}}) +  \sum_{i = 1}^{K} \mu_i \Omega_i \overline{\mathbf{H}}_i + \mu_P \mathbf{I}_N,  \label{lemma1-3}
\end{align}
where $\tilde{\mathbf{A}}(\theta_{\mathrm{B}}) = \mathbf{a}(\theta_{\mathrm{B}}) \mathbf{a}^H(\theta_{\mathrm{B}})$

Note that, for positive semidefinite matrices $\mathbf{L}_R$, $\mathbf{C}_R$, $\mathbf{L}_k$, $\mathbf{W}_k$, \eqref{lemma1-1d} is equivalent to $\mathbf{L}_R \mathbf{C}_R = 0$ and $\mathbf{L}_k \mathbf{W}_k = 0, \forall k $. 

Then, the equality $\mathbf{L}_R \mathbf{C}_R = 0$ means that the column space of $\mathbf{C}_R$ is  the null space of $\mathbf{L}_R$. Because, for any vector $\mathbf{x} = [x_1, x_2, \cdots, x_N]$,  $\mathbf{L}_R \mathbf{C}_R \mathbf{x} = \mathbf{L}_R \mathbf{y} = 0$, where  $\mathbf{y}  = \mathbf{C}_R \mathbf{x}$ is the null space of $\mathbf{L}_R$, and $\mathbf{C}_R \mathbf{x}$ is the column space of $\mathbf{C}_R$. 
In other hands, we have $\mathrm{Col}(\mathbf{C}_R) = \mathrm{Nul} ( \mathbf{L}_R )$, i.e., $\mathrm{rank} (\mathbf{C}_R ) = \mathrm{dim}( \mathrm{Col}(\mathbf{C}_R) ) = \mathrm{dim}(\mathrm{Nul} ( \mathbf{L}_R ))$,
where $\mathrm{Col}$ represents the column space of the matrix, $\mathrm{Nul} $ denotes the null space of the matrix, $\mathrm{dim}$ is the dimension of the subspace. 
Hence, by the rank–nullity theorem, we have $\mathrm{rank} (\mathbf{L}_R ) + \mathrm{rank} (\mathbf{C}_R ) = \mathrm{rank} (\mathbf{L}_R ) + \mathrm{dim}(\mathrm{Nul} ( \mathbf{L}_R )) = N$. 
Similarly, we can prove $\mathrm{rank} (\mathbf{L}_k ) + \mathrm{rank} (\mathbf{W}_k ) = N, \forall k$.

In the following, we need to prove $\mathrm{rank} (\tilde{\mathbf{W}}_k) = 1, \forall k$ and $\tilde{\mathbf{W}}_R = 0$. 
We discuss two cases: the first case is when $ \mu_k = 0 $ for all $ k $; the second case is when there exists at least one $ \mu_k > 0 $.

For the first case, according to complimentary conditions, \eqref{3-7b} is strictly less than $0$ at the optimal point.
Then, we have 
\begin{align}
    & \mathbf{L}_k = \mathbf{L}_R = - \tilde{\mathbf{A}}(\theta_{\mathrm{B}})  + \mu_P \mathbf{I}_N.  \label{lemma1-4} 
\end{align}

With $\mathbf{L}_k \succeq 0, \, \forall k$, $\mathbf{L}_R \succeq 0$, and rank 1 property of $\tilde{\mathbf{A}}(\theta_{\mathrm{B}})$, these two equalities imply that $\mu_P = \lambda_{A}$, $\mathrm{rank} (\mathbf{L}_k ) = \mathrm{rank} (\mathbf{L}_R ) = N-1$, where $\lambda_{A}$ is the eigenvalue of $\tilde{\mathbf{A}}(\theta_{\mathrm{B}})$. 
This further indicates that $ \mathbf{C}_R $ and $ \mathbf{W}_k $ lie in the subspace spanned by the eigenvector $\mathbf{u}$ corresponding to the unique non-zero eigenvalue $\lambda_{A}$ of $ \tilde{\mathbf{A}}(\theta_{\mathrm{B}}) $, which means all the beamformers of the BS point to the IRS. 
Note that, to guarantee the communication rate constraint, it must have $\mathrm{rank}(\mathbf{W}_k) = 1, \forall k$. 
Below, we can prove by contradiction that $ \mathbf{C}_R $ must be 0.
We firstly assume that $\mathrm{rank}(\overline{\mathbf{C}}_R) = 1, \overline{\mathbf{C}}_R = c \mathbf{u} \mathbf{u}^H$. 
For every optimal solution $\overline{\mathbf{W}}_k = \omega_k  \mathbf{u} \mathbf{u}^H$ and $\overline{\mathbf{C}}_R$, we can always construct a rank 1 solution $\breve{\mathbf{W}}_k = \overline{\mathbf{W}}_k + \iota_k \overline{\mathbf{C}}_R = (\omega_k + \iota_k c) \mathbf{u} \mathbf{u}^H, \sum_{k = 1}^{K} \iota_k = 1, \iota_k \geq 0$ achieving the optimality.
Here, $ \overline{\mathbf{W}}_k $ and $ \overline{\mathbf{C}}_R $ are spanned by the vector $ \mathbf{u} $, ensuring that $\breve{\mathbf{W}}_k $ also satisfies the rank-1 constraint.
Next, Problem \eqref{3-7} can be verified as
\begin{align}
    &\mathbf{a}^H(\theta_{\mathrm{B}}) \sum_{k = 1}^{K} \breve{\mathbf{W}}_k \mathbf{a}(\theta_{\mathrm{B}}) 
	= \mathbf{a}^H(\theta_{\mathrm{B}}) ( \sum_{k = 1}^{K} \overline{\mathbf{W}}_k +  \overline{\mathbf{C}}_R ) \mathbf{a}(\theta_{\mathrm{B}}), \label{lemma1-6} \\
    &\text{Tr}( \sum_{k = 1}^{K} \breve{\mathbf{W}}_k )  =  \text{Tr}( \sum_{k = 1}^{K} \overline{\mathbf{W}}_k  + \overline{\mathbf{C}}_R) \leq P_0, \label{lemma1-7} \\
    &\quad \Omega_k \text{Tr}( \overline{\mathbf{H}}_k  \sum_{k = 1}^{K} \breve{\mathbf{W}}_k )   - (1+\Omega_k) \text{Tr}( \overline{\mathbf{H}}_k  \breve{\mathbf{W}}_k ) + \Omega_k \sigma_{k}^2  \nonumber  \\
	& = \Omega_k \text{Tr} \big( \overline{\mathbf{H}}_k  ( \sum_{k = 1}^{K} \overline{\mathbf{W}}_k + \overline{\mathbf{C}}_R ) \big)    + \Omega_k \sigma_{k}^2 \nonumber \\
	&\quad  - (1+\Omega_k) \text{Tr} \big( \overline{\mathbf{H}}_k  (\overline{\mathbf{W}}_k + \iota_k \overline{\mathbf{C}}_R) \big)\nonumber  \\
	& \leq  \Omega_k \text{Tr} \big( \overline{\mathbf{H}}_k  ( \sum_{k = 1}^{K} \overline{\mathbf{W}}_k + \overline{\mathbf{C}}_R  ) \big)   - (1+\Omega_k) \text{Tr} \big( \overline{\mathbf{H}}_k  \overline{\mathbf{W}}_k   \big) \nonumber \\ 
	& \quad + \Omega_k \sigma_{k}^2 < 0, \quad \forall k.  \label{lemma1-8} 
\end{align}

Hence, there always exists rank 1 optimal solution $\mathbf{W}_k, \forall k$ and optimal solution $\mathbf{W}_R = 0$.

For the second case, where there exists at least one $ \mu_k > 0 $, we have
\begin{align}
    & \mathbf{L}_k \overset{(a)}{=} \mathbf{L}_R - (\mu_k + \mu_k \Omega_k) \overline{\mathbf{H}}_k, \quad \forall k, \label{lemma1-9} \\
    &  \text{Tr}( \mathbf{L}_k \mathbf{C}_R  ) =  \text{Tr}( \mathbf{L}_R \mathbf{C}_R  ) - 
    (\mu_k + \mu_k \Omega_k) \text{Tr}( \overline{\mathbf{H}}_k \mathbf{C}_R) \overset{(b)}{\geq} 0, \label{lemma1-10}
\end{align}
where $(a)$ follows from \eqref{lemma1-2} and \eqref{lemma1-3}, $(b)$ follows from $\mathbf{L}_k \succeq 0, \mathbf{C}_R \succeq 0$. 
Also, because $ \text{Tr}( \mathbf{L}_R \mathbf{C}_R) = 0, \text{Tr}( \overline{\mathbf{H}}_k \mathbf{C}_R) \geq 0$ and $\mu_k \geq 0 $, it follows that $ (\mu_k + \mu_k \Omega_k)\text{Tr}(\overline{\mathbf{H}}_k \mathbf{C}_R) = 0, \forall k $. 
Thus, using $ (\mu_k + \mu_k \Omega_k)\text{Tr}(\overline{\mathbf{H}}_k \mathbf{C}_R) = 0, \forall k $ and $\mathbf{L}_R \mathbf{C}_R = 0$, we can obtain
\begin{align}
   & \quad \big( \mu_P \mathbf{I}_N - \tilde{\mathbf{A}}(\theta_{\mathrm{B}}) \big) \mathbf{C}_R \nonumber \\
   &= \big( \mu_P \mathbf{I}_N - \tilde{\mathbf{A}}(\theta_{\mathrm{B}}) +  \sum_{i = 1}^{K} \mu_i \Omega_i \overline{\mathbf{H}}_i \big) \mathbf{C}_R = 0. \label{lemma1-11}
\end{align}

The equality above and $\mathrm{rank} (\mathbf{L}_R ) + \mathrm{rank} (\mathbf{C}_R ) \leq N$  imply that $\mathrm{rank} (\mathbf{C}_R) \leq 1, \mathrm{rank} \big( \mu_P \mathbf{I}_N - \tilde{\mathbf{A}}(\theta_{\mathrm{B}}) \big) \geq N - 1 $.
Assume that the dual variables greater than 0 is $ \mu_n, n \geq 1 $, then we have $ \overline{\mathbf{H}}_n \mathbf{C}_R = 0$. 
The situation where $ \mathbf{C}_R $ lies in the subspace pointing to the IRS and is simultaneously orthogonal to the subspaces of several communication equivalent channels, including both direct and reflected  links, is not possible. 
So, it must have $\mathbf{C}_R = 0$.

Next, we prove that $\mathrm{rank}(\mathbf{W}_k) = 1, \forall k$. To guarantee communication task, $\mathbf{W}_k \neq 0, \mathrm{rank}(\mathbf{W}_k) \geq 1, \forall k$.
So, with $\mathrm{rank} (\mathbf{L}_k ) + \mathrm{rank} (\mathbf{W}_k ) \leq N, \forall k$, it is easy to see that  $\mathrm{rank}(\mathbf{L}_k) \leq N - 1, \forall k$.
Then, we have 
\begin{align}
    \mathrm{rank}(\mathbf{L}_R) &= \mathrm{rank}\big( \mathbf{L}_k  + (\mu_k + \mu_k \Omega_k) \overline{\mathbf{H}}_k \big), \nonumber \\
    &  \leq \mathrm{rank}( \mathbf{L}_k) +  \mathrm{rank}\big(  (\mu_k + \mu_k \Omega_k) \overline{\mathbf{H}}_k \big), \quad \forall k. \label{lemma1-12}
\end{align}
Due to $\mathbf{C}_R = 0$ and $\mathrm{rank} (\mathbf{L}_R ) + \mathrm{rank} (\mathbf{C}_R ) = N$,  we have $\mathrm{rank} (\mathbf{L}_R) = N$, i.e.,
$\mathrm{rank}(\mathbf{L}_k) \geq N - 1, \forall k$.
So, there exists $\mathrm{rank}(\mathbf{W}_k) = 1, \forall k$, which completes the proof.


\section{ THE PROOF OF LEMMA \ref{lemma-2} \label{proof2}   }
First, the vector $\mathbf{g}_1$ can be transformed  as 
\begin{align}
	\mathbf{g}_1 
	&= -2   \text{Tr}(-\mathbf{M}_1)   \text{vec} ( \mathbf{e}^{(\tau)} (\mathbf{e}^{(\tau)})^H  ) - 2 \mathbf{M}_R    \text{vec} ( \mathbf{e}^{(\tau)} (\mathbf{e}^{(\tau)})^H  )  \nonumber \\
	&= 2 z_1  \text{vec} ( \mathbf{e}^{(\tau)} (\mathbf{e}^{(\tau)})^H  )  - 2   \sum_{p=1}^{N_r}  \beta_p^2 \mathbf{C}_p^H \overline{\mathbf{b}}(\theta_{\mathrm{I}}) \overline{\mathbf{b}}^H(\theta_{\mathrm{I}})  \nonumber \\
	& \quad \text{vec} \big(  \mathrm{diag}(\mathbf{b} (\theta_p^r)  ) \mathbf{e}^{(\tau)} (\mathbf{e}^{(\tau)})^H  {\mathrm{diag}(\mathbf{b}^* (\theta_p^r)  )}^T \big)  \nonumber \\
	&= 2 z_1  \text{vec} ( \mathbf{e}^{(\tau)} (\mathbf{e}^{(\tau)})^H  )  -
	2  \sum_{p=1}^{N_r}  \beta_p^2 \text{vec} (\mathbf{M}_3), \label{4-8}
\end{align}
where the second and third equalities are derived from the equation $\text{vec}(\mathbf{A} \mathbf{B} \mathbf{C}) = (  \mathbf{C}^T \otimes \mathbf{A} ) \text{vec}(\mathbf{B}) $ because  $\mathbf{C}_p$ and $\overline{\mathbf{b}}(\theta_{\mathrm{I}})$
are the Kronecker products of two matrices or vectors.

Then, by undoing  the vectorization in \eqref{4-8}, matrix $\mathbf{G}_1$ can be expressed as 
\begin{align}
	\mathbf{G}_1 =	2 z_1 \mathbf{e}^{(\tau)} (\mathbf{e}^{(\tau)})^H  - 2  \sum_{p=1}^{N_r}  \beta_p^2 \mathbf{M}_3. \label{4-9}
\end{align}

Hence, the proof is completed.

\section{ THE PROOF OF LEMMA \ref{lemma-3} \label{proof3}   }
 First, with \eqref{3-1}, sensing MI can be transformed as  \eqref{3-2}. 
    Then, we only need to derive an upper bound on \eqref{3-4a}.
    Let $\Psi(\mathbf{W})$ and $\Phi(\mathbf{E})$ denote $\overline{\mathbf{a}}^H(\theta_{\mathrm{B}}) \widetilde{\mathbf{W}}^H \widetilde{\mathbf{W}} \overline{\mathbf{a}}(\theta_{\mathrm{B}})$ and $\overline{\mathbf{b}}^H(\theta_{\mathrm{I}}) \mathbf{J} \mathbf{R}_{R} \mathbf{J}^H  \overline{\mathbf{b}}(\theta_{\mathrm{I}})$, respectively.
    Note that $\Psi(\mathbf{W})$ and $\Phi(\mathbf{E})$ are two independent scalars, each related to $\mathbf{W}$ and $\mathbf{E}$, respectively. 
    Next, we provide the upper bounds for these two terms.
    
    First, $\Psi(\mathbf{W})$ can be written as 
    \begin{align}
        \Psi(\mathbf{W}) &=  N \mathbf{a}^H(\theta_{\mathrm{B}}) \mathbf{W} \mathbf{W}^H \mathbf{a}(\theta_{\mathrm{B}}) 
        \nonumber  \\
        &= N \sum_{i=1}^{K+N} \mathbf{a}^H(\theta_{\mathrm{B}}) \mathbf{w}_i \mathbf{w}_i^H \mathbf{a}(\theta_{\mathrm{B}}) \nonumber  \\
        & = N \sum_{i=1}^{K+N} {| \mathbf{a}^H(\theta_{\mathrm{B}}) \mathbf{w}_i |}^2 \nonumber  \\
        & \leq  N \sum_{i=1}^{K+N} {|| \mathbf{a}(\theta_{\mathrm{B}})  ||}^2 {|| \mathbf{w}_i ||}^2 \label{lemma3-2} \\
        & = N^3 \sum_{i=1}^{K+N}  |\nu_i|^2, \label{lemma3-3}
    \end{align}
    where \eqref{lemma3-2} follows from the Cauchy-Schwarz inequality, and the equality holds when $\mathbf{w}_i = \nu_i \mathbf{a}(\theta_{\mathrm{B}}) $.
    
    Let $\mathbf{w}_i = \nu_i \mathbf{a}(\theta_{\mathrm{B}}) $.  The transmit power budget can be rewritten as
    \begin{align}
        \mathrm{Tr} (\mathbf{W} \mathbf{W}^H)  &= \sum_{i=1}^{K+N} |\nu_i|^2  \mathrm{Tr} \big( \mathbf{a}(\theta_{\mathrm{B}}) \mathbf{a}^H(\theta_{\mathrm{B}}) \big) \nonumber \\
        &= N \sum_{i=1}^{K+N} |\nu_i|^2 \leq P_0. \label{lemma3-4}
    \end{align}

    With \eqref{lemma3-3} and \eqref{lemma3-4}, we have $\Psi(\mathbf{W}) \leq N^2 P_0 $.

   Then, $\Phi(\mathbf{E})$ can be recast as 
   \begin{align}
       \Phi(\mathbf{E}) & = \beta^2 \big(\mathbf{b}^T (\theta_\mathrm{I}) \otimes \mathbf{b}^H (\theta_\mathrm{I})  \big) \big( \mathbf{E}^T \otimes \mathbf{E}^H \big)  \big(  \mathbf{b}^* (\theta^r)    \otimes \mathbf{b} (\theta^r) \big)  \nonumber \\
       & \quad \big(  \mathbf{b}^T (\theta^r)    \otimes \mathbf{b}^H (\theta^r) \big)  \big( \mathbf{E}^* \otimes \mathbf{E} \big)  \big(\mathbf{b}^* (\theta_\mathrm{I}) \otimes \mathbf{b} (\theta_\mathrm{I})  \big) \nonumber \\
       & = \beta^2 {\big( \mathbf{b}^T (\theta_\mathrm{I}) \mathbf{E}^T \mathbf{b}^* (\theta^r)    \mathbf{b}^H (\theta_\mathrm{I}) \mathbf{E}^H \mathbf{b} (\theta^r)  \big)}^2 \nonumber \\
       & = \beta^2  {\big( \mathbf{e}^H \mathrm{diag}( \mathbf{b}(\theta_\mathrm{I})  )^H   \mathbf{b} (\theta^r) \mathbf{b}^H (\theta^r)   \mathrm{diag}( \mathbf{b}(\theta_\mathrm{I})  ) \mathbf{e} \big)}^2 \nonumber \\
       & \leq \beta^2 \tilde{\lambda}^2 M^2,  \label{lemma3-5}
   \end{align}
    where  \eqref{lemma3-5} follows from $\mathbf{b} (\theta^r) \mathbf{b}^H (\theta^r) \preceq \tilde{\lambda} \mathbf{I}_{M}  $ and $ |e_m|^2 = 1, \forall m$.

    By combining the upper bounds for $\Psi(\mathbf{W})$ and $\Phi(\mathbf{E})$, we can derive the upper bound on sensing MI, i.e., \eqref{lemma3-1}, which completes the proof.

\bibliographystyle{IEEEtran}
\bibliography{ref}

\end{document}